\documentclass[12pt]{iopart}
\usepackage{graphicx}
\usepackage[caption=false]{subfig}
\usepackage{epstopdf}
\usepackage{amsfonts}

\begin{document}

   \title{The hodograph method for relativistic Coulomb systems}

   \author{Uri Ben-Ya'acov}

   \address{School of Engineering, Kinneret Academic College on
   the Sea of Galilee, \\   D.N. Emek Ha'Yarden 15132, Israel}

   \ead{uriby@kinneret.ac.il}

\vskip 2.0cm

\begin{abstract}
Relativistic Coulomb systems are studied in velocity space, prompted by the fact that the study of Newtonian Kepler/Coulomb systems in velocity space provides a method much simpler (and more elegant) than the familiar analytic solutions in ordinary space. The key for the simplicity and elegance of the velocity-space method is the linearity of the velocity equation, which is a unique feature of $1/r$ interactions for Newtonian and relativistic systems alike, allowing relatively simple analytic discussion with coherent geometrical interpretations. Relativistic velocity space is a 3-D hyperboloid ($H^3$) embedded in a 3+1 pseudo-Euclidean space. The orbits in velocity space for the various types of possible trajectories are discussed, accompanied with illustrations.
\end{abstract}



\vskip30pt

\section{Introduction}

An `hodograph' is the orbit in velocity space that corresponds to the trajectory of a particle in ordinary space, traversed by the tip of the velocity vectors when these are drawn all starting from the same point, the origin of velocity space. It was introduced by Hamilton \cite{Hamilton1847} as an essential tool in proving, using {\it geometrical} relations rather that analytical, that Kepler's 1st law follows from Newton's law of universal attraction, the laws of mechanics and Kepler's 2nd law. The hodographs of Keplerian motion -- indeed of all classical motion under $1/r$ potentials -- have the very special property that they are full circles or circular arcs for all possible trajectories, and the geometrical proof asserts the connection between these circles and the spatial trajectories being conic sections.

Hamilton's proposition was soon after elaborated by Maxwell \cite{Maxwell}. However, for some unclear reason (perhaps because the geometrical approach became less and less familiar to students) Hamilton's method and the hodograph are hardly mentioned in standard Mechanics textbooks, even such prominent ones as Goldstein's \cite{Goldstein.etal2000} or Landau \& Lifshitz' \cite{LLMech}. Keplerian motion is discussed there only analytically and only via the trajectory in ordinary space, although, for instance, Goldstein certainly knew of Hamilton's method \cite{Goldstein76}. So much so, that when Feynman prepared what came to be known as his `lost lecture' \cite{Goodstein96} he re-developed Hamilton's method anew, apparently unaware of the work of his predecessors. Since then there appear in the literature from time to time re-iterations and some extensions of Hamilton's method (many times also attributed to Feynman) for Kepler/Coulomb (KC) systems, mainly from a pedagogical point of view complementing the bits missing in textbooks \cite{Milnor1983,Sivardiere1992,GonVilla.etal,Butikov2000,Derbes2001,KowenMathur2003,Munoz2003,Carinena.etal2016}. It is also noted that hodographs became useful tools in dynamical systems other than KC \cite{Sivardiere1992,Apostolatos2003}.

However, even analytically, using the hodograph to solve Newtonian KC systems is much simpler than the analytic solution in ordinary space \cite{Milnor1983,Munoz2003}. The unique feature of KC systems, which allows this simplicity and elegance, is the {\it linearity} of the hodograph equation (see \eref{eq: veqangKC}), a direct consequence of the $1/r$ nature of the potential. Although the relativistic trajectories are much more complicated than the Newtonian ones, the virtue of linearity of the hodograph equations persists into the relativistic regime (equations \eref{eq: polangeq} \& \eref{eq: uoeqang}).

Besides an interest in relativistic applications of the hodograph method \textit{per se}, our interest in the hodograph solution for relativistic Coulomb systems also stems from an interest in the relativistic EM 2-body system, which so far doesn't have a satisfactory solution in the special-relativistic non-quantum realm \cite{Stephas1978}. The success of using the hodograph method with Newtonian KC systems gives rise to the hope that with its application to relativistic EM systems a general solution may be advanced. It is therefore also in this context that the present work has been performed.

Newtonian KC 2-body systems may be reduced to 1-body systems by transition to the centre-of-mass (CM) reference frame. Such a simple and direct procedure is impossible, in general, for relativistic 2-body systems because the interaction is not instantaneous. It is therefore convenient to start with Coulomb systems, which may be regarded as the limit of EM 2-body systems when one of the charges is much heavier than the other, modeling hydrogen-like atoms or Rutherford scattering. Such relativistic Coulomb systems were discussed only few times in the literature (see, \textit{e.g.}, \cite{LLF39}; also \cite{Torkelsson1998,Boyer2004,Stahlhofen2005a} and references therein), almost always using only standard analytic methods to determine the particle's trajectory in ordinary space.

The purpose of the present work is therefore to study the dynamics of a relativistic charged particle in Coulomb field in velocity space, aiming to obtain, possibly with some simplicity and elegance, new and further insights into the dynamics of such systems.

A brief but illustrated summary of our results is published elsewhere \cite{IARD2016}. The full presentation splits here into two parts. The present article brings the theory of the relativistic hodographs, presenting and discussing the various solutions. It starts with reviewing the classical hodograph method (\Sref{sec: hodKC}) and presentation of the relativistic velocity space (\Sref{sec: RVS}). Then the relativistic hodograph equations are discussed (\Sref{sec: hodrelC}) and general properties of the hodographs are deduced thereof (\Sref{sec: Genprop}). These are followed by explicit discussion, accompanied with exemplary illustrations, of all the possible solutions of the hodograph equations (\Sref{sec: ellarge} \& \ref{sec: excrel}). The special internal symmetry of these systems is then discusses in a companion article \cite{Hamsym}.

{\it Notation}. The convention $c=1$ is used throughout, unless specified otherwise. Events in Minkowski space-time are $x^\mu = \left(x^0,x^1,x^2,x^3\right)$, with metric tensor $g_{\mu\nu} = {\rm diag} \left(-1,1,1,1\right) \, , \, \mu,\nu = 0,1,2,3$. For any 4-vectors $a^\mu = (a^0,\vec a)$ and $b^\mu = (b^0,\vec b)$, their inner product is then $a \cdot b = -a^0 b^0 + \vec a \cdot \vec b$.

\vskip20pt

\section{The hodograph in classical Kepler/Coulomb systems} \label{sec: hodKC}

We start with reviewing the essentials of the classical KC analytic hodograph solution. Further information and discussions may be found in various publications \cite{Milnor1983,Sivardiere1992,GonVilla.etal,Butikov2000,Derbes2001,KowenMathur2003,Munoz2003,Carinena.etal2016}.

Considering the classical KC equation of motion
 \begin{equation} \label{eq: peqmot}
 \frac{d \vec p}{d t} = \frac{\kappa}{r^3} \vec r \, ,
 \end{equation}
it is convenient to assume the $x,y$ coordinates for the plane of motion, with the conserved angular momentum $\vec \ell  = \vec r \times \vec p = \ell \hat z$ perpendicular to it. We also use polar coordinates $(r,\theta)$ with the polar-planar unit vectors
 \begin{equation} \label{eq: poluvec}
 \hat r = \cos \theta \hat x + \sin \theta \hat y  \quad ,  \quad  \hat \theta  =  - \sin \theta \hat x + \cos \theta \hat y
 \end{equation}
satisfying
 \begin{equation} \label{eq: poluvec2}
 \hat r = - \hat z \times \hat \theta = - \frac{d \hat\theta}{d \theta}  \quad , \quad  \hat \theta = \hat z \times \hat r = \frac{d \hat r}{d \theta} \, .
 \end{equation}

An essential aspect of the Hamilton-Feynman method is the change from $t$-dependence to $\theta$-dependence, employing Kepler's 2nd law which is the angular-momentum conservation law,
 \begin{equation} \label{eq: angmomKC}
 \ell = m r^2 \frac{d\theta}{dt} \, .
 \end{equation}
Then, with $\vec p = m \vec v$, the momentum equation \eref{eq: peqmot} becomes simply
\begin{equation}\label{eq: veqangKC}
 \frac{d \vec v}{d \theta} = \frac{\kappa}{\ell} \hat r \, ,
\end{equation}
and using the relation $\hat r = - d\hat\theta / d\theta$ the solution of \eref{eq: veqangKC} is immediate,
 \begin{equation} \label{eq: hodoN}
\vec v = \vec B_o - \frac{\kappa}{\ell} \hat \theta \, ,
 \end{equation}
with $\vec B_o$ some arbitrary constant vector. Clearly, the simplicity of equation \eref{eq: veqangKC} which leads to this solution is due to the $1/r^2$ force.

\begin{figure}
  \centering
  \subfloat[The hodograph as the circle $C_o(\ell)$. \label{fig:1a}] {\includegraphics[width=5cm]{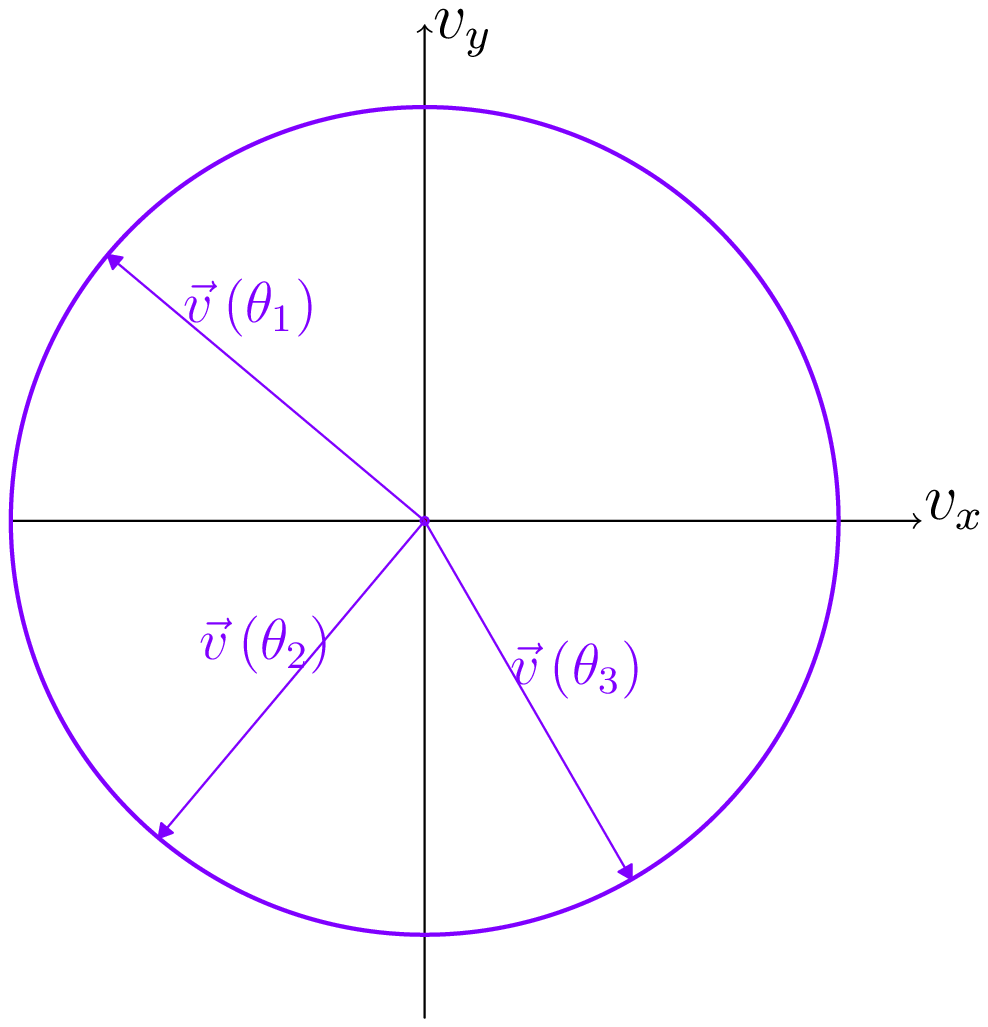}} \hskip40pt
  \subfloat[Spatial circular motion. \label{fig:1b}] {\includegraphics[width=5cm]{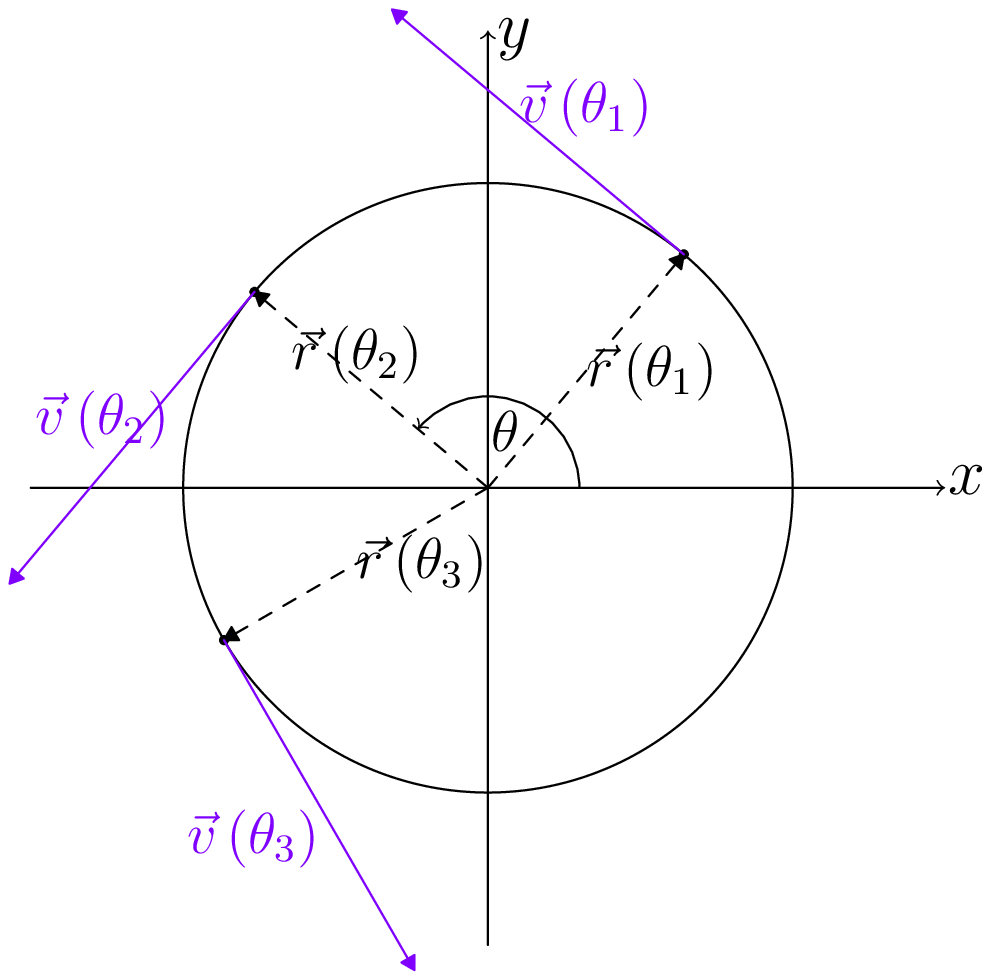}}
  \caption{\label{fig:1} Newtonian KC hodographs for minimum energy states : With spatial circular motion (right) the hodograph (left) is the canonical circle $C_o(\ell)$ centred at the origin of the velocity space.}
\end{figure}
\begin{figure}
  \centering
  \subfloat[The hodograph as the circle $C(\ell,E')$, displaced from the origin by $\vec B_o$. The velocity vectors are $\vec v (\theta) = \vec B_o + \vec v_o (\theta)$. \label{fig:2a}] {\includegraphics[width=5cm]{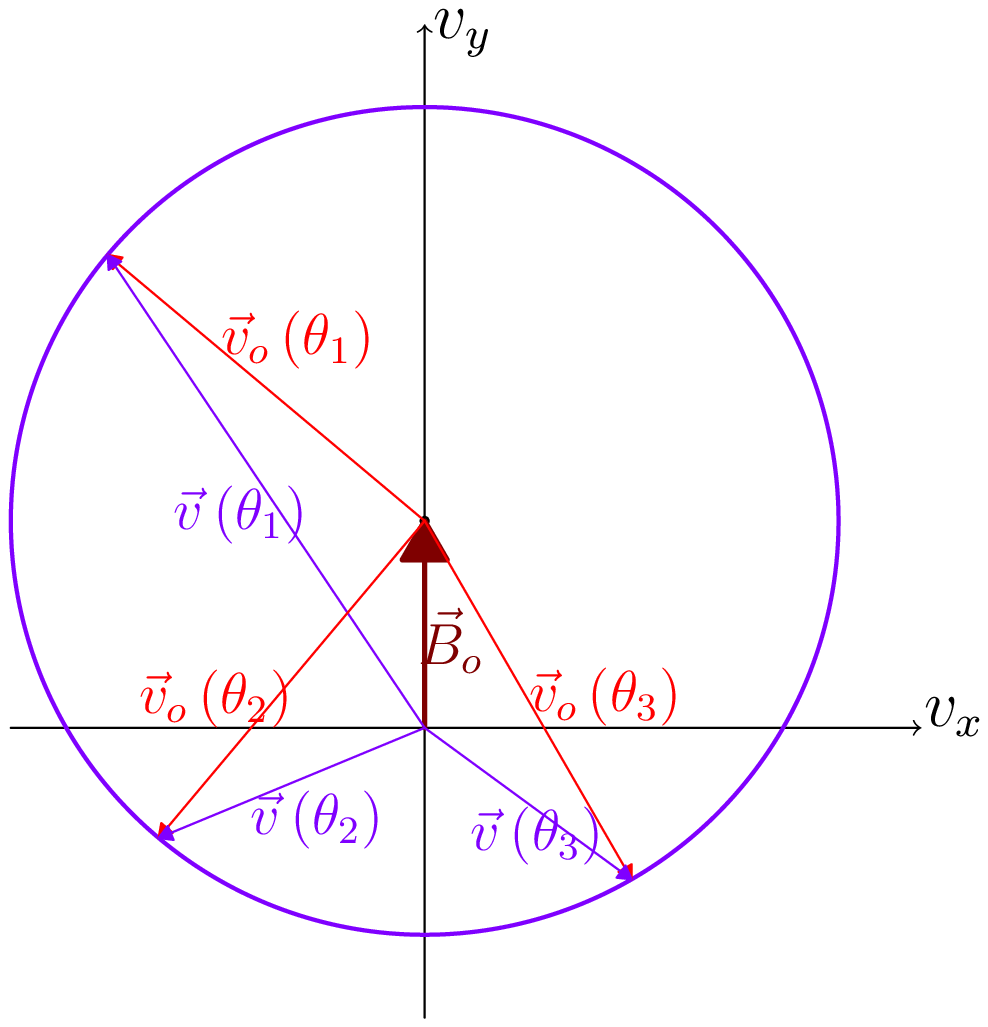}} \hskip40pt
  \subfloat[Spatial elliptic motion. \label{fig:2b}] {\includegraphics[width=5cm]{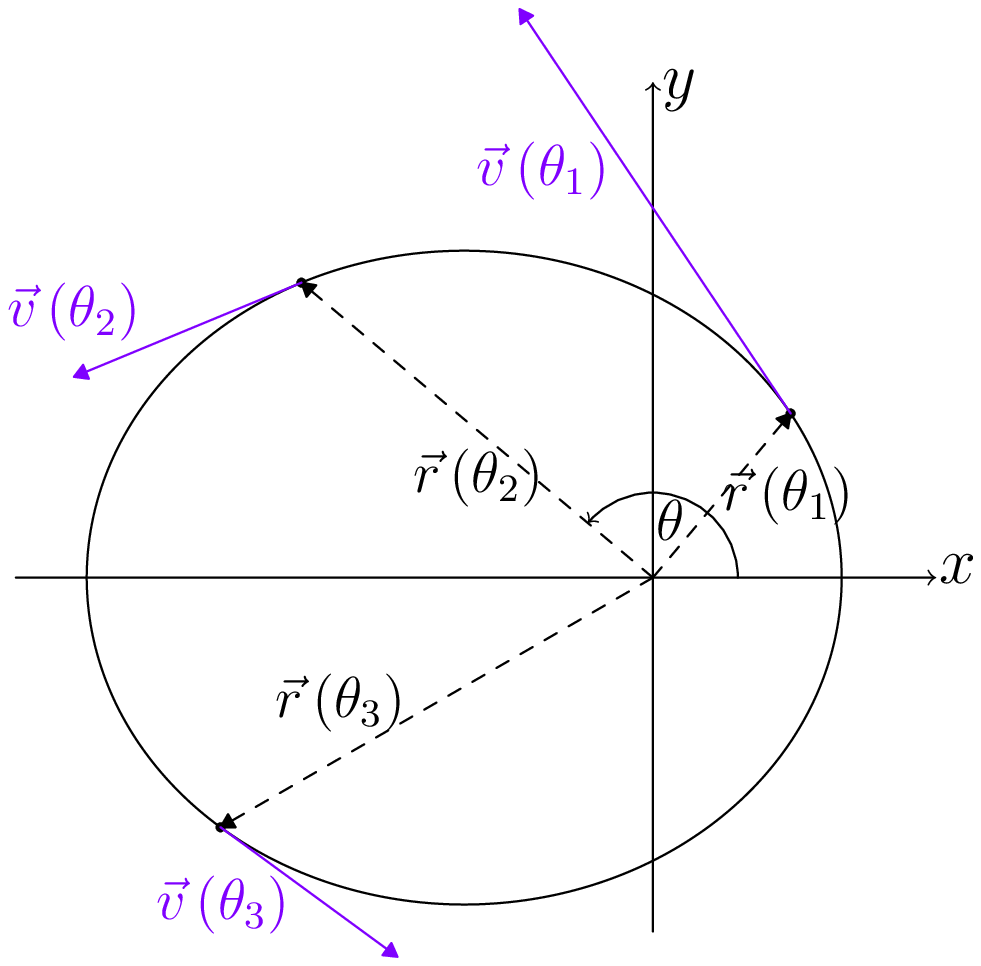}}
  \caption{\label{fig:2} Newtonian KC hodographs for general bound states : With spatial elliptic orbit (right) the hodograph (left) is the circle $C(\ell,E')$ centred at $\vec B_o$.}
\end{figure}

\begin{figure}
  \centering
  \subfloat[Attraction. \label{fig:3a}] {\includegraphics[width=10cm]{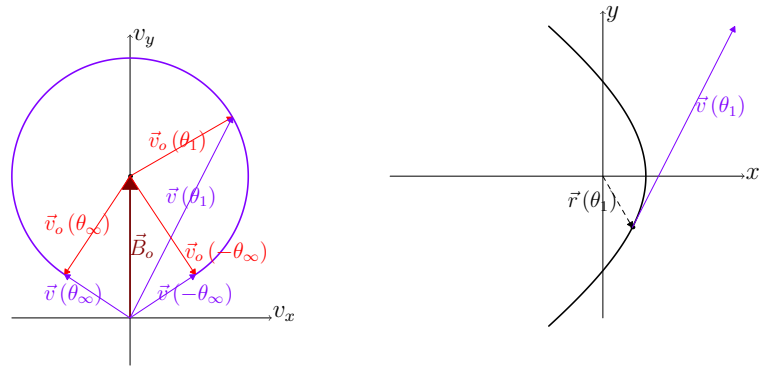}} \\
  \subfloat[Repulsion. \label{fig:3b}] {\includegraphics[width=10cm]{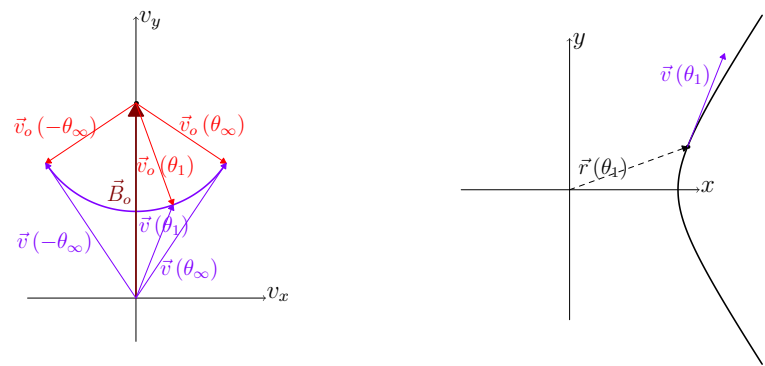}}
  \caption{\label{fig:3} Newtonian KC hodographs for unbound states : With spatial hyperbolic orbits (right) the hodographs (left) are finite circular arcs in the angular range $-\theta_\infty \le \theta \le \theta_\infty$, with radii $|\kappa|/\ell$ and centred at $\vec B_o$. The velocity vectors are $\vec v (\theta) = \vec B_o + \vec v_o (\theta)$ and are tangent to the arc at both ends.}
\end{figure}

The solution \eref{eq: hodoN} describes a circle (or at least a circular arc for unbound systems) in velocity space,
 \begin{equation} \label{eq: hodoN2}
 \left( \vec v - \vec B_o \right)^2 = \frac{\kappa^2}{\ell^2} \, ,
 \end{equation}
centred around $\vec B_o$ and with radius $|\kappa| / \ell$. Using the relation $v_\theta = r \dot\theta = \ell/(mr)$, the energy integral becomes, for total energy $E\,'$,
\begin{equation}\label{eq: energyintN}
 \frac{m {\vec v \,}^2}{2} + \frac{\kappa}{r} = \frac{m {\vec v \,}^2}{2} + \frac{m \kappa}{\ell} v_\theta = E\,'
\end{equation}
from which it is easily verified, substituting \eref{eq: hodoN} in \eref{eq: energyintN}, that
\begin{equation}\label{eq: poNewt}
 {B_o}^2 = \frac{2E\,'}{m} + \frac{\kappa^2}{\ell^2} \, ,
\end{equation}
For any given value of $\ell$, the case of minimal energy $E'$ is with $\vec B_o = 0$. The corresponding hodograph is a canonical circle $C_o(\ell)$ drawn by the tips of the vectors $\vec v_o (\theta) = -\left(\kappa/\ell\right) \hat \theta$ centred at the origin of velocity space, with circular spatial motion (\Fref{fig:1}) \footnote{The illustrations in Figures \ref{fig:1}, \ref{fig:2} \& \ref{fig:3} share a common colour code : Velocity vectors $\vec v(\theta)$ are purple, the Hamilton vector $\vec B_o$ is brown, and the vectors $\vec v_o(\theta)$ are in red. These colours cannot be seen in the printed version, and the reader is advised to use the on-line or PDF versions.}. With increasing energy $\vec B_o$ becomes non-zero, and the corresponding hodographs \eref{eq: hodoN} remain circles $C(\ell,E')$ with the same radius $|\kappa|/\ell$ but now centred at $\vec B_o$ with elliptic spatial orbits (\Fref{fig:2}). For unbound systems the hodographs reduce the infinite spatial trajectory into a finite circular arc -- another merit of the hodograph method -- whose centre is again at $\vec B_o$ and with the same radius $|\kappa|/\ell$ (\Fref{fig:3}). The endpoints of the arc correspond to $v_\theta = 0 \Leftrightarrow r \rightarrow \infty$, with $\vec v$ tangent to the hodograph.

The (constant) vector $\vec B_o$ is known as the {\it Hamilton vector}. Its magnitude $B_o$, depending on the energy via \eref{eq: poNewt}, determines the relative position of the vector : for $E\,' <,=,>0$ the vector $\vec B_o$ lies within the hodograph, touches it or exceeds it, corresponding to elliptic, parabolic or hyperbolic spatial trajectories, respectively (see, \textit{e.g.}, \cite{Sivardiere1992,GonVilla.etal,Butikov2000}). In either case, the physical interpretation is that the Hamilton vector $\vec B_o$ transforms the hodographs between configurations that correspond to different values of the energy, while keeping the same value of the angular momentum.

\vskip20pt

\section{The relativistic velocity space}\label{sec: RVS}

The relativistic velocity space (RVS) is the space of all future-directed time-like unit 4-vectors, and its usage has the virtue of transforming and displaying kinematical and dynamical properties in a purely geometrical manner. Since there is no explicit dependence on time, different points in it may correspond either to the velocity states of different particles in some instantaneous reference frame, or the velocity state of a particle in different instances along its world-line.

To the author's best knowledge, the most thorough account of the RVS is by Rhodes and Semon \cite{RhodesSemon2004}. They point at an exercise contained already in the 1951 English edition of Landau \& Lifshitz' ``Classical theory of fields" \cite{LLF5} as the first time the idea of RVS appeared in English (for an account of the earlier history of the RVS see a comment by Criado and Alamo \cite{CriadoAlamo2001}). While the Landau \& Lifshitz' exercise dealt only with the metric properties of the RVS, later applications of the RVS \cite{RhodesSemon2004,Urbantke1990,Aravind1997} focus on geometrical derivation of the Thomas-Wigner rotation.

So far, the association of geometry to the RVS used the spatial velocity $\vec v$ for the RVS coordinates, and the metric properties derived from the Lorentz formulae for relativistic velocity addition \cite{LLF5,CriadoAlamo2001}. Instead, we use here the relativistic velocity 4-vector $u^\mu = \left(\gamma\left(v\right), \gamma\left(v\right) \vec v\right)$ with the RVS defined as
\begin{equation}\label{eq: Vrel}
 {\cal V}_{\rm rel} \equiv \left\{ u^\mu = \left(u^0 , \vec u \right) | u^0 = \sqrt{1 + \vec u\, ^2} \right\}
\end{equation}
This is a 3-D unit hyperboloid ($H^3$) embedded in a 4-D pseudo-Euclidian space
\begin{equation}\label{eq: E13}
 E^{(1,3)} = \left\{ w^\mu = \left( w^0,\vec w \right) \in \mathbb{R}^4 | g_{\mu\nu} = {\rm diag} \left(-1,1,1,1\right) \right\}
\end{equation}
Also, for any $u^\mu \in {\cal V}_{\rm rel}$, let ${\cal T}_u\left({\cal V}_{\rm rel}\right) \subset E^{(1,3)}$ be the hyperplane tangent to ${\cal V}_{\rm rel}$ at $u^\mu$. Any vector $A^\mu \in {\cal T}_u\left({\cal V}_{\rm rel}\right)$ satisfies $A^2 > 0 , \, A \cdot u = 0$.

Using $u^\mu$ as the coordinates of the RVS allows to relate the geometrical properties of the hyperbolic space with relativistic kinematics and dynamics in a Lorentz-covariant manner. The geometrical properties of the RVS then follow straight-forward. Further properties of the RVS are discussed in the companion article \cite{Hamsym}.

\vskip20pt

\section{The hodograph equation for relativistic Coulomb systems} \label{sec: hodrelC}

A relativistic Coulomb system consists of a point particle with mass $m$ whose dynamics is determined by the Hamiltonian \cite{LLF39}
 \begin{equation} \label{eq: Ham}
 H\left(\vec r, \vec p \right) = \sqrt {{\vec p \,}^2 + m^2}  + \frac{\kappa}{r}
 \end{equation}
with the equation of motion identical in form to the Newtonian one \eref{eq: peqmot}
 \begin{equation} \label{eq: eqmot}
 \frac{d \vec p}{d t} = \frac{\kappa}{r^3} \vec r \, .
 \end{equation}
The total energy $E = H\left(\vec r, \vec p \right)$ and angular momentum $\vec \ell  = \vec r \times \vec p$ are conserved, the latter satisfying
 \begin{equation} \label{eq: angmom}
 \ell = \left( E - \frac{\kappa}{r} \right) r^2 \frac{d \theta}{d t}
 \end{equation}
as the relativistic generalization of the Newtonian angular-momentum conservation law \eref{eq: angmomKC}.

Solving the momentum equation \eref{eq: eqmot} in the velocity space provides an elegant and simple solution for the motion in a relativistic Coulomb system. The key to the hodograph or velocity space picture is the angular momentum conservation law \eref{eq: angmom}, the relativistic counterpart of Kepler's 2nd law. It allows, as above, transition to $\theta$ as the hodograph parameter. The (spatial) linear momentum of the particle is $\vec p = m\vec u$, so the momentum equation \eref{eq: eqmot} becomes
\begin{equation}\label{eq: vecueqang}
 \frac{d \vec u}{d \theta} = \frac{\kappa}{\ell} u^0 \hat r \, .
\end{equation}
Using the polar representation $\vec u = {u_r}\hat r + {u_\theta } \hat\theta$, the polar-angular equations derived from \eref{eq: vecueqang} are
\begin{equation}\label{eq: polangeq}
 \frac{d u_\theta}{d \theta} = -u_r \quad , \quad \frac{d u_r}{d \theta} = \frac{\kappa}{\ell} u^0 + u_\theta \, ,
\end{equation}
complemented by the $u^0$-equation also derived from \eref{eq: vecueqang},
\begin{equation}\label{eq: uoeqang}
 \frac{d u^0}{d \theta} = \frac{\vec u}{u^0} \cdot \frac{d \vec u}{d \theta} = \frac{\kappa}{\ell} \vec u \cdot \hat r = \frac{\kappa}{\ell} u_r \, .
\end{equation}
Equations \eref{eq: vecueqang} (or, alternatively, \eref{eq: polangeq}) and \eref{eq: uoeqang} constitute the relativistic hodograph equations.

The relation
\begin{equation}\label{eq: uthetar}
 u_\theta = \frac{\ell}{m r} \, ,
\end{equation}
another consequence of \eref{eq: angmom}, allows transition from spatial dependencies to velocity dependencies. In particular, the conserved energy \eref{eq: Ham} is
\begin{equation}\label{eq: energyint}
 m u^0 + \frac{\kappa}{r} = E \, ,
\end{equation}
so using \eref{eq: uthetar}, the energy integral \eref{eq: energyint} becomes
\begin{equation}\label{eq: energyint2}
 u^0 + \frac{\kappa}{\ell} u_\theta = \frac{E}{m}
\end{equation}
which may also be recognized as an immediate integral of equations \eref{eq: polangeq} and \eref{eq: uoeqang}.

The virtue of equations \eref{eq: polangeq}, \eref{eq: uoeqang} and \eref{eq: energyint2} is their linearity with constant coefficients in the polar representation, a unique feature of the $1/r$ interaction. Therefore, explicit solutions providing the hodographs and the corresponding spatial trajectories are quite immediate to get. However, there is much information and insight that may be obtained from just studying the equations themselves. We start with this study in the following section, deferring the explicit solutions to \sref{sec: ellarge} and \ref{sec: excrel}.

\vskip20pt

\section{General properties of the relativistic hodographs}\label{sec: Genprop}

\subsection{The hodograph equations in the embedding space}\label{sec: hodeqE}

Recalling that the velocity space ${\cal V}_{\rm rel}$ is embedded in a psudo-Euclidean space $E^{(1,3)}$, we note that the hodograph equations \eref{eq: vecueqang} and \eref{eq: uoeqang} may be combined and considered for arbitrary orbits $w^\mu \left( \theta \right) = \left(w^0, \vec w \right)$ in $E^{(1,3)}$, not necessarily confined to $\mathcal{V}_{\rm rel}$, as
\begin{equation}\label{eq: weqgen}
 \frac{d w^\mu}{d \theta} = \frac{\kappa}{\ell} {\Omega^\mu}_\nu w^\nu
\end{equation}
with
\begin{equation}\label{eq: Omega}
{\Omega^\mu}_\nu = \left(\begin{array}{*{20}{c}}
  0 & \vline & {\hat r} \\
\hline
  {\hat r} & \vline & 0
\end{array}
\right) = \left(\begin{array}{*{20}{c}}
  0 & \vline & {\cos \theta} & {\sin \theta} & 0 \\
\hline
  {\cos \theta} & \vline & 0 & 0 & 0 \\
  {\sin \theta} & \vline & 0 & 0 & 0 \\
  0 & \vline & 0 & 0 & 0
\end{array}
\right)
\end{equation}
Since $\Omega_{\mu\nu} = -\Omega_{\nu\mu}$, it immediately follows, as a general property of the hodograph equations, that if $w_1^\mu \left( \theta \right)$ and $w_2^\mu \left( \theta \right)$ are both orbits in $E^{(1,3)}$ that satisfy \eref{eq: weqgen} then the inner product $w_1 \cdot w_2$ is constant. Therefore, any solution of the hodograph equations is of constant magnitude, and their motion can only be rotational.

Since the hodograph equations in the polar representation \eref{eq: polangeq} \& \eref{eq: uoeqang} are with constant coefficients, it follows that \eref{eq: weqgen} generates rotation in $E^{(1,3)}$ relative to the polar coordinates. In order to identify the axis of rotation, we note that the 4-vector
\begin{equation}\label{eq: vodef}
 v_o^\mu = \left( 1, \vec v_o \right) = \left( 1, - \frac{\kappa}{\ell} \hat\theta \right)
\end{equation}
is a particular solution of \eref{eq: weqgen} which is constant relative to the polar coordinates with components $\left(v_{or}, v_{o\theta}\right) = \left(0,-\kappa/\ell \right)$ and $v_o \cdot w = {\rm const.}$ for any solution $w^\mu$ of \eref{eq: weqgen}. $v_o^\mu$ is therefore identified as the axis vector, relative to the polar system, of the pseudo-rotation generated by \eref{eq: weqgen} in $E^{(1,3)}$.

The physical significance of the axis vector $v_o^\mu$ is that the energy integral \eref{eq: energyint2} is identified, for any hodograph $u^\mu$ in $\mathcal{V}_{\rm rel}$, as the constant $u$-component along $v_o^\mu$,
\begin{equation}\label{eq: energyint3}
 \frac{E}{m} = -u \cdot v_o \, .
\end{equation}
Its spatial part is the classical 3-vector $\vec v_o = - \left( \kappa/\ell \right) \hat \theta$ which generates the base canonical circle $C_o \left(\ell\right)$ whose shifting by the Hamilton vector $\vec B_o$ produces the (Newtonian) hodograph. $v_o^\mu$ is therefore the $E^{(1,3)}$ extension of $\vec v_o$.

\subsection{General properties of the relativistic hodographs with relations to the spatial trajectories}

It is evident from equations \eref{eq: polangeq} - \eref{eq: energyint2} and the foregoing discussion that the nature of the solution, both for the spatial trajectory and the hodograph, depends on the dimensionless parameters $E/m$ and $\kappa/\ell$. Some further insights thus ensue :
\begin{enumerate}
 \item {The relation $u_\theta = \ell / m r$ \eref{eq: uthetar} implies $u_\theta \ge 0$. Since $u^0 \ge 1$ it follows from the energy integral \eref{eq: energyint2} that
\begin{equation}\label{eq: utcond}
 \frac{\kappa }{\ell} u_\theta \le \frac{E}{m} -1
\end{equation}
Therefore, if $E < m$ then $\kappa$ can only be negative (attraction), and for $\kappa > 0$ (repulsion) necessarily $E \ge m$.}
  \item {The relation $u_\theta = \ell / m r$ also implies that $u_\theta = 0$ (corresponding to $r \to \infty$) is possible only for unbound systems. The relation $u^0 = \sqrt{1 + {u_r}^2 + {u_\theta}^2}$ together with the energy constraint \eref{eq: energyint2} yield the relation for the polar components,
\begin{equation}\label{eq: urutrel}
 {u_r}^2 + \left( 1 - \frac{\kappa^2}{\ell^2} \right) {u_\theta}^2 + \frac{2\kappa E}{m\ell} u_\theta = \frac{E^2}{m^2} - 1 \, .
\end{equation}
(This looks like an equation of conic sections in velocity space, but the variables here are the polar components and not Cartesian ones, so the actual orbit is much more complicated.) Then it follows from \eref{eq: urutrel} that for the limiting states
\[
 u_{r,\infty}^2 = \frac{E^2}{m^2} - 1
\]
and therefore necessarily $E \ge m$ for unbound systems.}
  \item {The limiting states $u_\theta = 0$ for $r \to \infty$ define corresponding angles $\theta_{\infty 1}$ and $\theta_{\infty 2}$, so that an infinite spatial trajectory becomes a finite segment in velocity space with $\theta$ in the range $\theta_{\infty 1} \le \theta  \le \theta_{\infty 2}$. Near $\theta_{\infty 1}$ the particle's motion is towards the centre of force, therefore there $u_{r,\infty 1} = u_r \left(\theta_{\infty 1} \right) = - \sqrt {E^2/m^2 - 1}$, while near $\theta_{\infty 2}$ the particle recedes from the centre of force and there $u_{r,\infty 2} = u_r \left(\theta_{\infty 2} \right) = \sqrt {E^2/m^2 - 1}$.}
  \item {Since in both endpoints $\vec u_\infty = u_{r,\infty} \hat r\left( \theta_\infty \right)$, result of $u_{\theta,\infty} = 0$, it follows that $\left( d\vec u / d\theta \right)_\infty$ is parallel to $\vec u_\infty$ in both endpoints. This means that in the projection of the hodograph on the $u_x$-$u_y$ plane, the vector $\vec u_\infty$ is tangent to the hodograph. It is the same characteristic of the endpoints as in the Newtonian case.}
  \item {The relativistic velocity space, reduced to $u_z = 0$ (the plane of motion), may be parameterized as
\begin{equation}\label{eq: umupar}
 \fl \hskip20pt u^\mu = \left( \cosh \eta , \sinh \eta \left( \cos \phi \hat x + \sin \phi \hat y \right) \right) \quad , \quad \eta \ge 0 \, , \, 0 \le \phi < 2\pi
\end{equation}
$u^0 = \cosh \eta$ identifies $\eta$ as the \textit{rapidity}. $\phi$, the azimuthal angle in velocity space, is different from $\theta$, which is the azimuthal angle in ordinary space and in velocity space serves only as the orbit parameter. At the endpoints $u^\mu_\infty = \left( E / m , u_{r,\infty} \hat r_\infty \right)$, therefore comparison with \eref{eq: umupar} yields
\begin{equation}\label{eq: etaphitheta}
 \eta_\infty = \cosh^{-1} \left(\frac{E}{m}\right) \quad , \quad  \phi_{\infty 1} = \theta_{\infty 1} + \pi  \quad , \quad \phi_{\infty 2} = \theta_{\infty 2}
\end{equation}
}
\end{enumerate}

\subsection{The hodograph dependence on the ratio $\left| \kappa \right| / \ell$}

Further insight requires explicit solution of the hodograph equations. The relativistic hodograph equations are \eref{eq: polangeq} and \eref{eq: uoeqang}, constrained with \eref{eq: energyint2}. These equations may be uncoupled by computing second derivatives, yielding, for the polar components,
\begin{equation}\label{eq: uAeq}
 \frac{d^2 u_A}{d \theta^2} + \left( 1 - \frac{\kappa^2}{\ell^2} \right) u_A = \frac{E}{m} v_{o,A}
\end{equation}
for $A=0,r,\theta$. This will serve now as the basic equation from which the solutions, in all cases, will be deduced.

It is clear that the nature of the solution depends on the coefficient of $u_A$ in \eref{eq: uAeq}, namely on the ratio $|\kappa| / \ell$. This is directly related with the magnitude of the axis vector \eref{eq: vodef}
\begin{equation}\label{eq: vomag}
{v_o}^2 = \frac{\kappa^2}{\ell^2} - 1 \quad \begin{array}{*{20}{c}}
 \nearrow \\
 \to \\
 \searrow
\end{array} \quad \begin{array}{*{20}{c}}
{\ell > |\kappa|}&{\Leftrightarrow}&{v_o^\mu \quad {\textrm {time-like}}}\\
{\ell = |\kappa|}&{\Leftrightarrow}&{v_o^\mu \quad {\textrm {light-like}}}\\
{\ell < |\kappa|}&{\Leftrightarrow}&{v_o^\mu \quad {\textrm {space-like}}}
\end{array}
\end{equation}
$|\kappa| / \ell$ is the magnitude of the spatial velocity on the basis circle formed by $v_o^\mu$, $C_o(\ell) = \left\{v_o^\mu(\theta)\right\}$, whose projection on the $w_x$-$w_y$($w^0$)-plane is the minimal energy circle in the Newtonian limit $\left\{\vec v_o(\theta)\right\}$ (see \Fref{fig:1}; the same notation $C_o(\ell)$ is used for both circles for convenience). Only for $\ell > |\kappa|$ are the velocities on $C_o(\ell)$ subluminal. Therefore, the relativistic requirement that particles' velocities cannot reach the velocity of light necessarily implies that the relativistic solution can have a non-relativistic limit only for $v_o^\mu$ time-like ($\ell > |\kappa|$). These cases with Newtonian limit are discussed in the following section. The cases for $\ell \le |\kappa|$ don't have a Newtonian counterpart, and are discussed in \sref{sec: excrel}.

\begin{figure}[h]
  \centering
  \includegraphics[width=6.5cm]{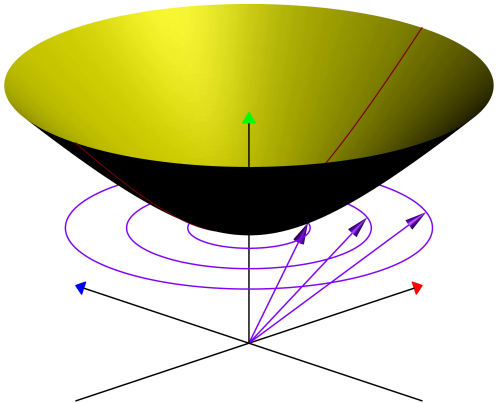}
  \begin{minipage}[b]{250pt}
  \caption{\label{fig:4} \textbf{The velocity space hyperboloid} ${\cal V}_{\rm rel}$. The axis vector $v_o^\mu$ and the basis circle $C_o\left(\ell\right)$ that it traverses, relative to ${\cal V}_{\rm rel}$, for the 3 possible cases :
  (i) left arrow -- $v_o^\mu$ time-like ($\ell > |\kappa|$);
  (ii) middle arrow -- $v_o^\mu$ light-like ($\ell = |\kappa|$);
  (iii) right arrow -- $v_o^\mu$ space-like ($\ell < |\kappa|$).}
  \end{minipage}
\end{figure}

A note regarding the hodograph illustrations in the following : Since the relativistic hodographs are 3-D curves, each of the cases discussed is illustrated with 3 projected views of a segment of the hodograph -- from above, from the side (horizontal) and an oblique-diagonal view -- accompanied by an illustration of the corresponding spatial orbit. All these illustrations share a common colour code : The thick red line is the hodograph itself, and in the side and oblique views the orange line is a cross-section of the velocity space hyperboloid ${\cal V}_{\rm{rel}}$ in the $\left(u_x = u_y\right)\textrm{-}u^0$ plane; the axes with red, blue and green arrowhead correspond, respectively, to $u_x$, $u_y$ and $u^0$ (the $u_z$ axis suppressed). These colours cannot be seen in the printed version, and the reader is advised to use the on-line or PDF versions. The scales are different according to case.

\vskip20pt

\section{Hodographs with Newtonian limit ($\ell > |\kappa|$)} \label{sec: ellarge}

Assuming $\ell > |\kappa|$ and introducing the notation
\begin{equation}\label{eq: beta}
 \beta = \sqrt {1 - \frac{\kappa^2}{\ell^2}} \, ,
\end{equation}
the combined solution to equation \eref{eq: uAeq} is found
\begin{equation}\label{eq: usol}
\begin{eqalign}
 { \vec u &=  B_o \sin \left( \beta \theta - \varphi \right) \hat r + \left[ - \frac{\kappa E}{m \ell \beta^2} + \frac{B_o}{\beta} \cos \left( \beta \theta - \varphi \right) \right] \hat \theta \\
 u^0 &= \frac{E}{m} - \frac{\kappa}{\ell} u_\theta = \frac{E}{m \beta^2} - \frac{\kappa B_o}{\beta\ell} \cos \left( \beta \theta - \varphi \right) }
\end{eqalign}
\end{equation}
where $\varphi$ is an arbitrary shift angle and $B_o$ is determined by substituting \eref{eq: usol} in \eref{eq: urutrel},
\begin{equation}\label{eq: Bo}
 B_o = \sqrt {\frac{E^2}{\beta^2 m^2} - 1} \, .
\end{equation}
The lower bound for the energy is therefore $E \ge \beta m$.

Using the identification $u_\theta = \ell / m r$ \eref{eq: uthetar}, the equation of the orbit in ordinary space is immediately obtained from the hodograph solution \eref{eq: usol} as
\begin{equation}\label{eq: r-orbit}
 \frac{1}{r} = \frac{m u_\theta}{\ell} = - \frac{\kappa E}{\ell^2 \beta^2} + \frac{m B_o}{\beta \ell} \cos \left( \beta \theta - \varphi \right) \, ,
 \end{equation}
recognized as a conic section for $\beta\theta$ as the polar angle, thus rotating conic section relative to $\theta$.

\subsection{The Newtonian limit}

In the Newtonian limit $u^\mu \to \left(1,\vec v\right)$, and, introducing $c$ (light velocity) explicitly, the dimensionless ratio $\ell / (|\kappa| c) \to 0$. Then the hodograph equation \eref{eq: vecueqang} reduces to \eref{eq: veqangKC} with its solution \eref{eq: hodoN}. Indeed, in the Newtonian limit the relativistic solution \eref{eq: usol} reduces to
 \begin{equation} \label{eq: vecvN}
\vec v = B_o \sin \left( \theta - \varphi \right) \hat r + \left[ - \frac{\kappa}{\ell} + B_o \cos \left( \theta - \varphi \right) \right] \hat \theta = B_o \hat \varphi  - \frac{\kappa}{\ell} \hat \theta
 \end{equation}
where $\hat \varphi = - \sin \varphi \hat x + \cos \varphi \hat y$, thus identifying the vector $\vec B_o = B_o \hat \varphi$ as the classical Hamilton vector. Similarly, denoting $E\,' = E - m{c^2}$, \eref{eq: poNewt} is easily recognized as the Newtonian limit of \eref{eq: Bo}.

\subsection{Bound states}

\begin{figure}
  \centering
  \subfloat[Hodograph -- side view \label{fig:5a}]{\includegraphics[width=4.5cm]{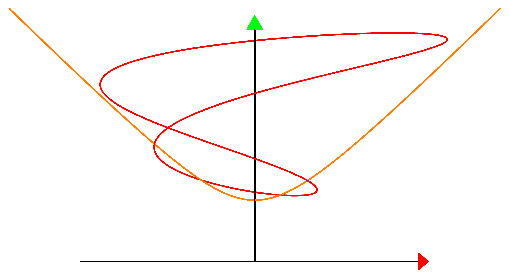}}\qquad
  \subfloat[Hodograph -- oblique view \label{fig:5b}]{\includegraphics[width=4.5cm]{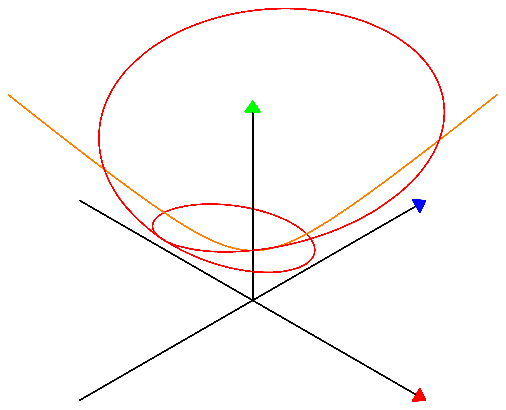}}\\
  \subfloat[Hodograph -- above view \label{fig:5c}]{\includegraphics[width=4.5cm]{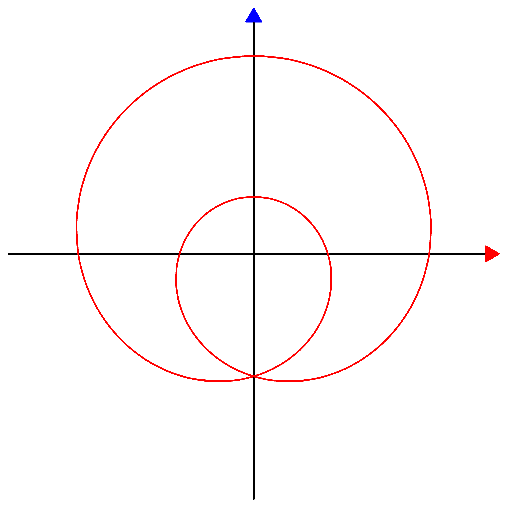}}\qquad
  \subfloat[Spatial orbit \label{fig:5d}]{\includegraphics[width=4.5cm]{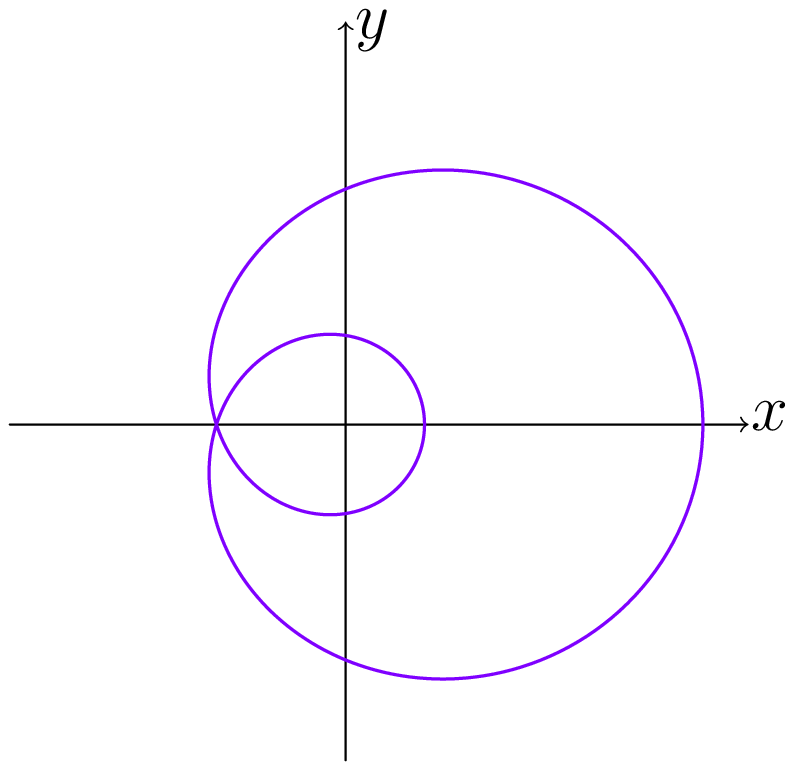}}
  \caption{\label{fig:5} Hodograph and spatial orbit for the bound state for $v_o^\mu$ time-like ($E = 0.6m, \kappa/\ell = -\sqrt{3}/2$). The trajectory is closed (periodical) because in this particular case $\beta = 1/2$.}
\end{figure}

Returning to the relativistic case, it is easy to verify from \eref{eq: usol}, using \eref{eq: Bo}, that for $E < m$ $u_\theta$ cannot vanish, corresponding to bound states. Extrema are obtained when $u_r = 0$, namely $\beta \theta = \varphi + n\pi \, , \, n  \in \mathbb{Z}$, with
 \begin{equation} \label{eq: utboundex}
 \frac{|\kappa|E}{m\ell \beta^2} - \frac{B_o}{\beta} \le u_\theta \le \frac{|\kappa|E}{m\ell \beta^2} + \frac{B_o}{\beta}
 \end{equation}
Corresponding lower and upper bounds for $r$ are found from $u_\theta = \ell/mr$,
 \begin{equation} \label{eq: rminmax}
 \frac{\left| \kappa \right| E - m \ell \beta B_o }{m^2 - E^2} \le r \le \frac{|\kappa| E + m \ell \beta B_o }{m^2 - E^2} \, .
 \end{equation}
Since $\beta \ne 1$ the hodograph is not periodic $2\pi$ in $\theta$, therefore not circular and the spatial trajectories are rotating ellipses. Periodicity is obtained only when $\beta$ is a rational number. An exemplary hodograph, using the values $E = 0.6m, \, \kappa/\ell = -\sqrt{3}/2 \, (\beta = 0.5)$ together with the corresponding spatial orbit obtained via \eref{eq: r-orbit} is demonstrated in \Fref{fig:5}.

\subsection{Unbound states}

For unbound systems, with $E \ge m$, $u_\theta$ vanishes (corresponding to $r \to \infty$) for the angles satisfying
 \begin{equation} \label{eq: rinfty}
 \cos \left( \beta \theta_\infty - \varphi \right) = \frac{\kappa E}{m\ell \beta B_o} \, ,
 \end{equation}
determining the angular range of the spatial trajectory. The infinite spatial trajectory becomes a finite segment in velocity space, just as in the corresponding Newtonian cases. The endpoints of the hodograph are at
 \begin{equation} \label{eq: thetainfty}
 \theta_\infty = \pm \frac{1}{\beta} \cos^{-1} \left( \frac{\kappa E}{m \ell \beta B_o} \right) + \frac{\varphi}{\beta} = \cases{
 \frac{ \pm \psi_\infty + \varphi}{\beta} & $\kappa > 0$ \\
 \frac{ \pm \left(\pi - \psi_\infty \right) + \varphi}{\beta} & $\kappa < 0$ \\
}
 \end{equation}
with
 \begin{equation} \label{eq: cospsiinfty}
 \psi_\infty \equiv \cos^{-1} \left(\frac{|\kappa|E}{m \ell \beta B_o}\right)  \quad  \left(0 < \psi_\infty < \pi / 2\right)
 \end{equation}
For actual computations in the following it is also convenient to obtain from \eref{eq: cospsiinfty}
 \begin{equation} \label{eq: sinpsiinfty}
 \sin \psi_\infty = \frac{\sqrt {E^2 - m^2}}{m B_o}
 \end{equation}

To get more insight into the significance of these result let $\varphi = 0$ so that $\vec B_o = B_o \hat y$, and let us consider separately the cases of repulsion ($\kappa > 0$) and attraction ($\kappa < 0$).

\begin{figure}
  \centering
  \subfloat[Hodograph -- side view\label{fig:6a}]{\includegraphics[width=4.5cm]{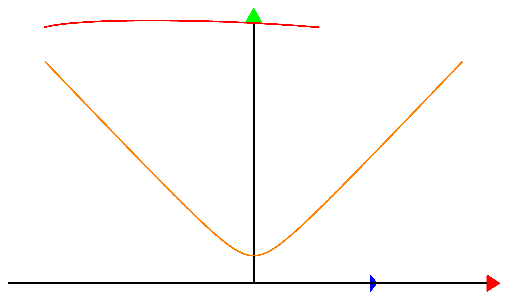}}\qquad
  \subfloat[Hodograph -- oblique view\label{fig:6b}]{\includegraphics[width=4.5cm]{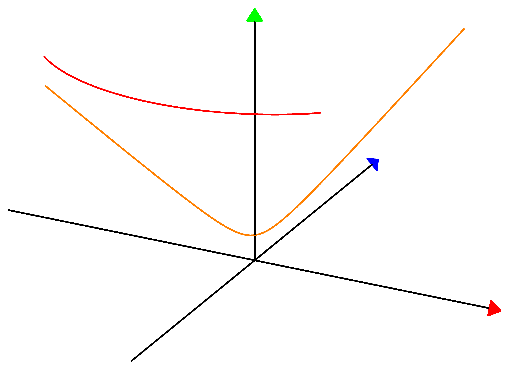}}\\
  \subfloat[Hodograph -- above view\label{fig:6c}]{\includegraphics[width=4.5cm]{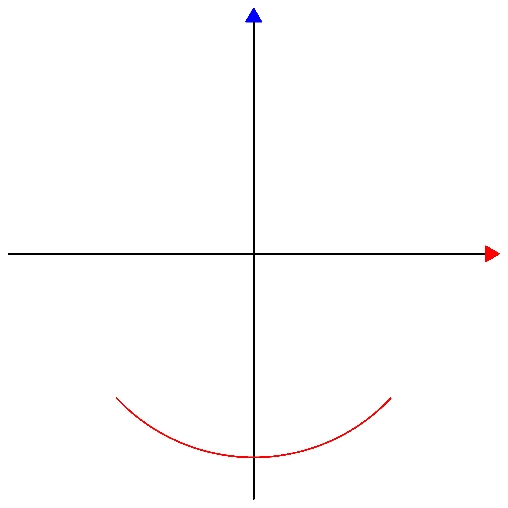}}\qquad
  \subfloat[Spatial orbit \label{fig:6d}]{\includegraphics[width=4.5cm]{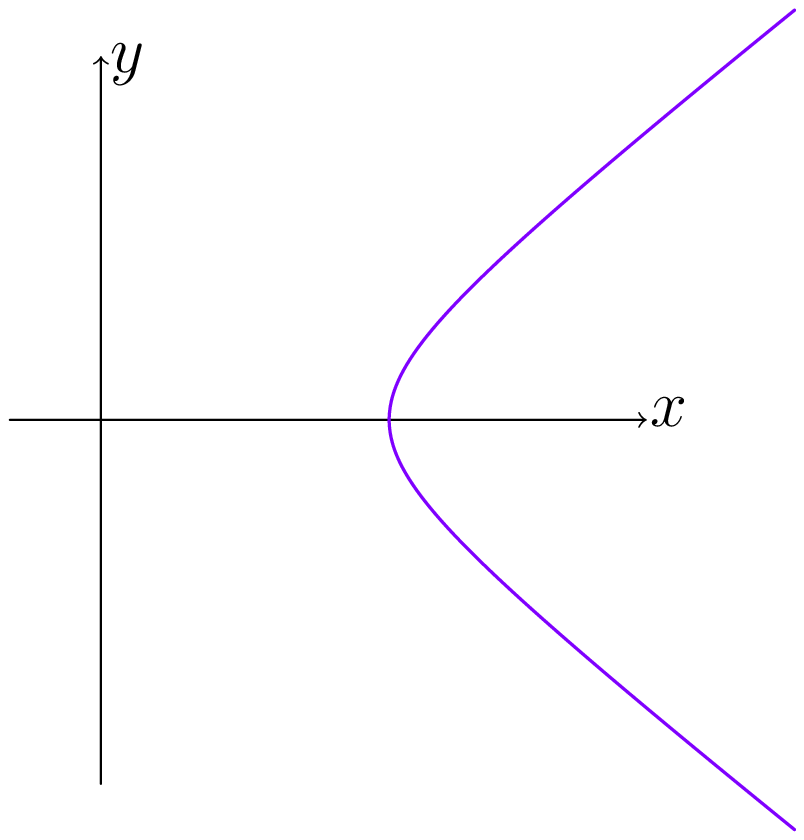}}
  \caption{\label{fig:6} Hodograph and spatial orbit for the repulsion unbound state for $v_o^\mu$ time-like ($E = 1.25m, \kappa/\ell = \sqrt{3}/2$). The infinite spatial trajectory becomes a finite segment in velocity space.}
\end{figure}

\subsubsection{Unbound states -- repulsion}

For $\kappa > 0$ the endpoints are at
\[
 \theta_{\infty 1} = - \frac{\psi_\infty}{\beta}  \quad  ,  \quad  \theta_{\infty 2} = \frac{\psi_\infty}{\beta}
\]
In the nonrelativistic limit, where $\beta \to 1$, the bounds are, as is well-known, $\left| {\theta_\infty} \right| < \pi / 2$, but in the ultra-relativistic case, with $\beta \to 0$, it follows from \eref{eq: sinpsiinfty} that the bounds become
\[
\left| {\theta_\infty} \right| = \frac{\psi_\infty}{\beta} = \frac{\psi_\infty}{\sin \psi_\infty} \sqrt {\frac{E^2 - m^2}{E^2 - \beta^2 m^2}} \to \frac{ \sqrt {E^2 - m^2}}{E} < 1\left[ {\rm{rad}} \right]
\]
The relativistic effect here is therefore contraction of the $\theta$-range. In velocity space the corresponding bounds are, from \eref{eq: etaphitheta},
\[
 \phi _{\infty 1} = \theta_{\infty 1} + \pi  = \pi - \frac{\psi_\infty}{\beta}  \quad  ,  \quad  \phi_{\infty 2} = \theta_{\infty 2} = \frac{\psi_\infty}{\beta}
\]
$\phi_{\infty 1}$ is in the second quarter while $\phi_{\infty 2}$ is in the first, therefore in the course of motion $\phi : \, \phi_{\infty 1} \to \phi_{\infty 2}$ so that $d\phi / d\theta <0$, and relative to the angle $\phi$ the spatial velocity $\vec u$ is in retrograde, qualitatively the same as for the Newtonian case (see \Fref{fig:6}).

\begin{figure}
  \centering
  \subfloat[Hodograph -- side view \label{fig:7a}]{\includegraphics[width=4.5cm]{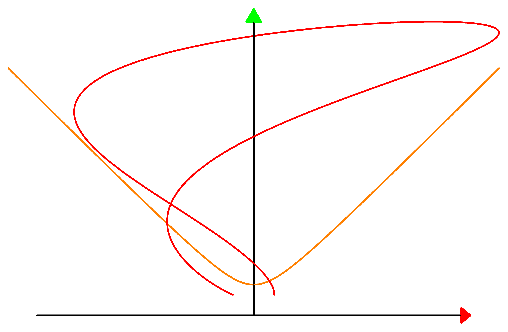}}\qquad
  \subfloat[Hodograph -- oblique view \label{fig:7b}]{\includegraphics[width=4.5cm]{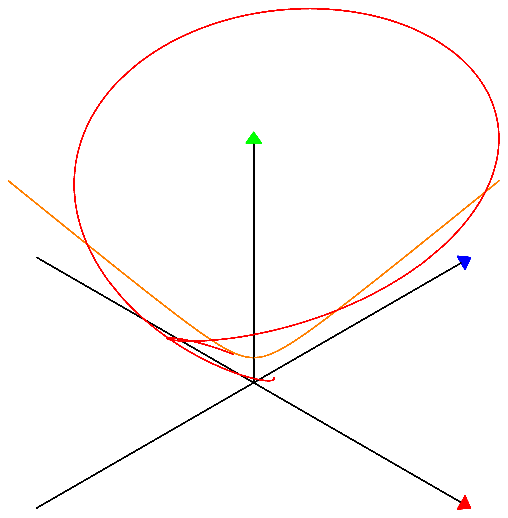}}\\
  \subfloat[Hodograph -- above view \label{fig:7c}]{\includegraphics[width=4.5cm]{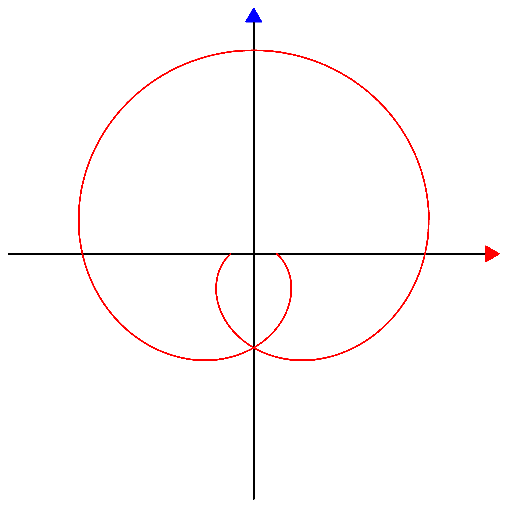}}\qquad
  \subfloat[Spatial orbit \label{fig:7d}]{\includegraphics[width=4.5cm]{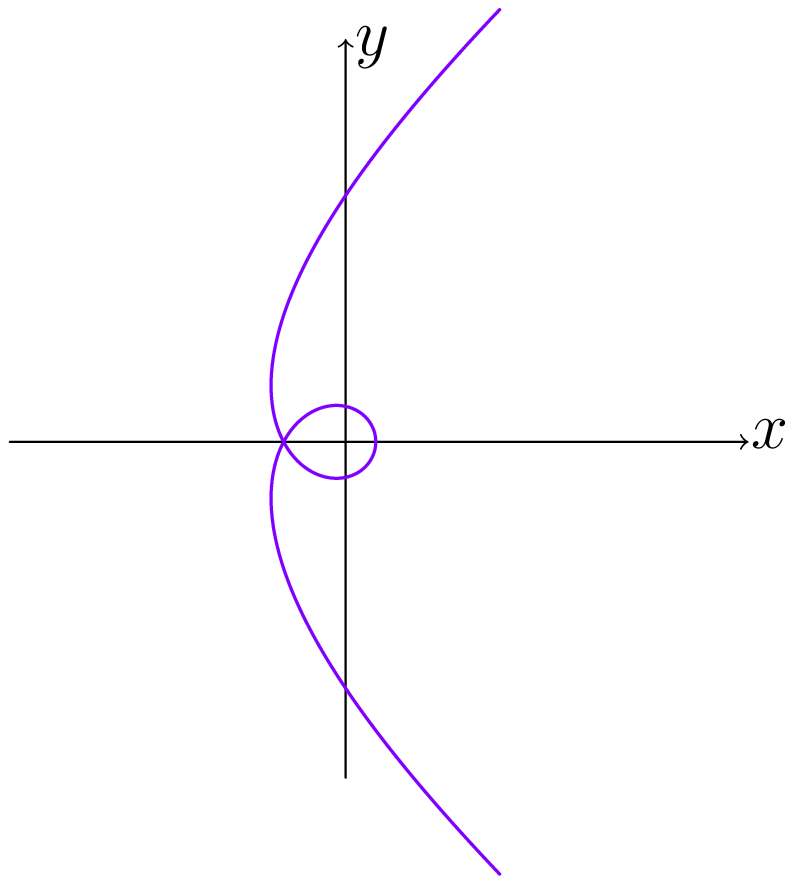}}
  \caption{\label{fig:7} Hodograph and spatial orbit for the attraction unbound state for $v_o^\mu$ time-like ($E = 1.25m, \kappa/\ell = -\sqrt{3}/2$).}
\end{figure}

\subsubsection{Unbound states -- attraction}

For $\kappa < 0$ the endpoints are at
\[
 \theta_{\infty 1} = - \frac{\pi - \psi_\infty}{\beta}  \quad  ,  \quad  \theta_{\infty 2} = \frac{\pi - \psi_\infty}{\beta}
\]
The inequality $\psi_\infty < \sin^{-1}\beta$ which follows from \eref{eq: Bo} and \eref{eq: sinpsiinfty} implies
 \begin{equation} \label{eq: tinftyconatt}
  \left| \theta_\infty \right| = \frac{\pi - \psi_\infty}{\beta} > \frac{\pi}{\beta} - \frac{\sin^{-1}\beta}{\beta}
 \end{equation}
Here there is no upper bound for $\left| \theta_\infty \right|$. In the nonrelativistic limit, with $\beta \to 1$, $\left| \theta_\infty \right| = \pi - \psi_\infty$, so the size of the $\theta$-range, $\theta_{\infty 2} - \theta_{\infty 1}$, doesn't exceed $2\pi$, and the hodograph is just an arc. But in the ultra-relativistic case with $\beta \to 0$ the lower bound of $\left| \theta_\infty \right|$ satisfies $\left| \theta_\infty \right|_{\min} \approx {\pi / \beta} - 1$, so that $\left| \theta_\infty \right| > \pi$, and the spatial particle's trajectory revolves around the centre of force at least once before running away to infinity. In fact, the rhs of \eref{eq: tinftyconatt} equals $\pi$ already for $\beta \approx 0.736$, corresponding to $\ell \approx 1.48|\kappa|
$.

Therefore, opposite to repulsion, in case of attraction the relativistic effect is to {\it enlarge} the $\theta$-range, in principle without limit, depending on $\beta$. The number of revolutions of the spatial trajectory around the centre of force is given by
 \begin{equation} \label{eq: nrev}
 N_{\rm{rev}} = \left[ \frac{\theta_{\infty 2} - \theta_{\infty 1}}{2\pi} \right] = \left[ \frac{2\pi - 2\psi_\infty}{2\pi \beta} \right] = \left[ \frac{\pi - \psi_\infty}{\pi \beta} \right]
 \end{equation}
where $\left[x\right]$ is the integer part of $x$. In velocity space the endpoints are at
 \begin{equation} \label{eq: phithetainf}
 \fl \hskip10pt \phi_{\infty 1} = \theta_{\infty 1} + \pi  = \pi - \frac{\pi - \psi_\infty}{\beta} = - \frac{\left(1 - \beta \right) \pi - \psi_\infty}{\beta}  \quad   ,  \quad  \phi_{\infty 2} = \theta_{\infty 2} = \frac{\pi - \psi_\infty}{\beta}
 \end{equation}
The angle $\phi$ is defined relative to the origin of velocity space, so the hodograph may also revolve around the origin, the number of revolutions being
 \begin{equation} \label{eq: nrev2}
 N_{\rm{rev}}' = \left[ \frac{\phi_{\infty 2} - \phi_{\infty 1}}{2\pi} \right] = \left[ \frac{\pi - 2\psi_\infty}{2\pi \beta} \right] = \left[ \frac{0.5\pi - \psi_\infty}{\pi \beta} \right]
 \end{equation}
Since the size of the $\theta$-range $\theta_{\infty 2} - \theta_{\infty 1}$ is always larger than the size of the $\phi$-range $\phi_{\infty 2} - \phi_{\infty 1}$ by $\pi$, then either $N_{\rm{rev}}' = N_{\rm{rev}}$ or $N_{\rm{rev}}' = N_{\rm{rev}} - 1$.

It should be noted, however, that since $-\pi + \psi_\infty \le \beta \theta \le \pi - \psi_\infty$, $u_r$ vanishes just once, for $\theta = 0$, with only one value for $r_{\rm min}$. The particle therefore spirals inwards with $r$ monotonically decreasing, reaches $r_{\rm min}$ and then spirals outwards with $r$ monotonically increasing.

As a numerical example, let $E = 1.25m$ and $\left| \kappa \right|/ \ell  = {\sqrt 3} / 2$ so that $\beta = 1/2$. Then $\psi_\infty \approx 0.106\pi$, and the endpoints are at
\[
\begin{array}{*{20}{c}}
{} & {\theta_{\infty 1}} & {\theta_{\infty 2}} & {\phi_{\infty 1}} & {\phi_{\infty 2}} \\
{\kappa  > 0 \, :} & {- 0.212\pi} & {0.212\pi} & {0.788\pi} & {0.212\pi} \\
{\kappa  < 0 \, :} & {- 1.788\pi} & {1.788\pi} & {- 0.788\pi} & {1.788\pi}
\end{array}
\]
For $\kappa < 0$ both the spatial trajectory and the hodograph make one complete revolution, with $N_{\rm{rev}} = \left[ 1.788 \right] = 1$ and $N_{\rm{rev}}' = \left[ 1.288 \right] = 1$ (see \Fref{fig:7}).

\vskip20pt

\section{Hodographs for exclusively relativistic cases} \label{sec: excrel}

The cases with $\ell \le |\kappa|$ don't have a nonrelativistic counterpart\footnote{As pointed out in \cite{LLF39} and \cite{Boyer2004}, a non-relativistic limit exists in these cases only for $\ell = 0$, but then the hodograph method doesn't work.}. First we consider separately the mathematically different solutions for $\ell < |\kappa|$ and $\ell = |\kappa|$, and then the structures of the hodographs and the particle's spatial trajectories are jointly discussed. The spatial trajectories in some of these cases with possible physical interpretation were discussed by Boyer \cite{Boyer2004}, with reference to a work by Darwin \cite{Darwin1913} which seems to be the first time these trajectories appeared in the literature.

\begin{figure}
  \centering
  \subfloat[Hodograph -- side view \label{fig:8a}]{\includegraphics[width=4.5cm]{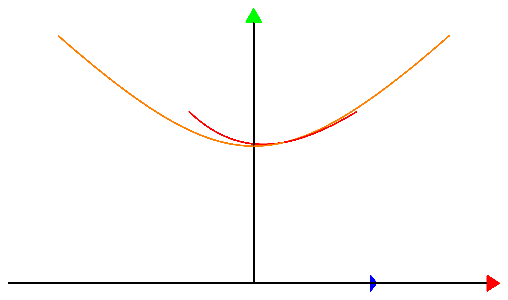}} \qquad
  \subfloat[Hodograph -- oblique view \label{fig:8b}]{\includegraphics[width=4.5cm]{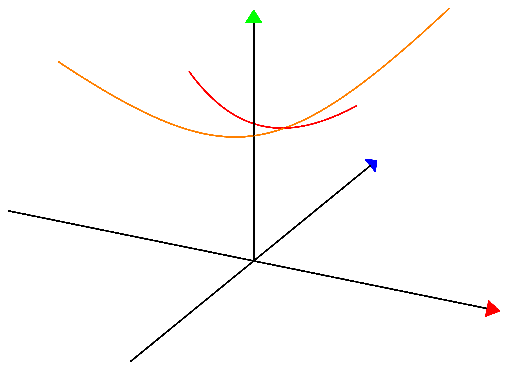}} \\
  \subfloat[Hodograph -- above view \label{fig:8c}]{\includegraphics[width=4.5cm]{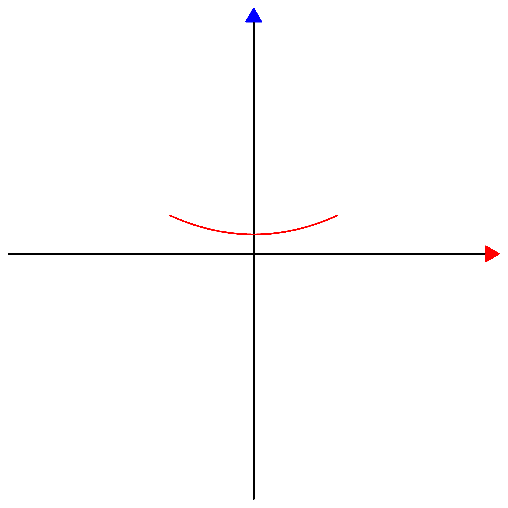}} \qquad
  \subfloat[Spatial orbit \label{fig:8d}]{\includegraphics[width=4.5cm]{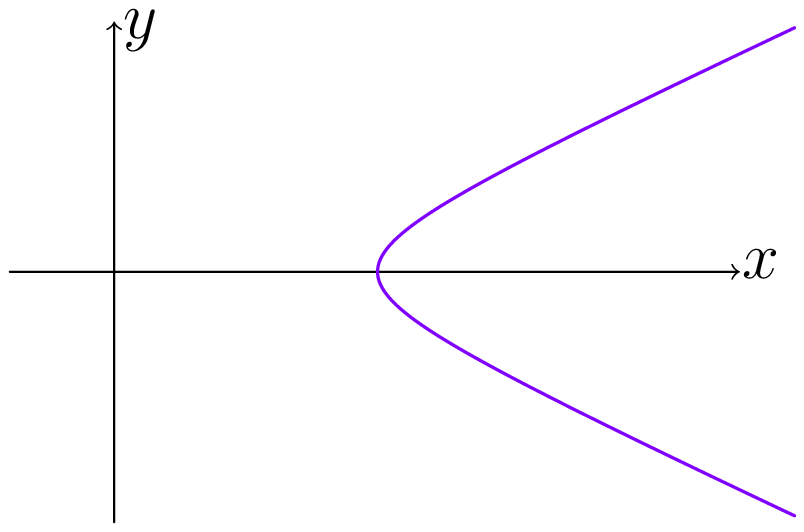}}
  \caption{\label{fig:8} Hodograph and spatial orbit for repulsion for $v_o^\mu$ space-like ($E = 1.25m, \kappa/\ell = 1.5$).}
\end{figure}

\subsection{Hodographs for $\ell < |\kappa|$} \label{sec: ellsmall}

Introducing the notation
\begin{equation}\label{eq: betabar}
 \bar\beta = \sqrt {\frac{\kappa^2}{\ell^2} - 1} \, ,
 \end{equation}
the solution of the hodograph differential equation \eref{eq: uAeq}, compliant with \eref{eq: energyint2} and \eref{eq: urutrel} and satisfying  $u^0 \ge 1$, is
\begin{equation}\label{eq: uklarge}
\begin{eqalign}
 {\vec u &= -\epsilon A_o \sinh \left[\bar\beta \left( \theta - \theta_o \right)\right] \hat r + \left\{ \frac{\kappa E}{m\ell \bar\beta^2} - \frac{\epsilon A_o}{\bar\beta} \cosh \left[\bar\beta \left( \theta - \theta_o \right)\right] \right\} \hat\theta \\
 u^0 &= - \frac{E}{m \bar\beta^2} + \frac{|\kappa| A_o}{\bar\beta \ell} \cosh \left[\bar\beta \left( \theta - \theta_o \right)\right] }
\end{eqalign}
\end{equation}
with the coefficient
\begin{equation}\label{eq: Ao}
 A_o = \sqrt {\frac{E^2}{m^2 \bar\beta^2} + 1} \, ,
 \end{equation}
$\epsilon = \rm{sign} \left(\kappa\right)$ and $\theta_o$ an arbitrary constant shift angle.

\begin{figure}
  \centering
  \subfloat[Hodograph -- side view\label{fig:9a}]{\includegraphics[width=4.5cm]{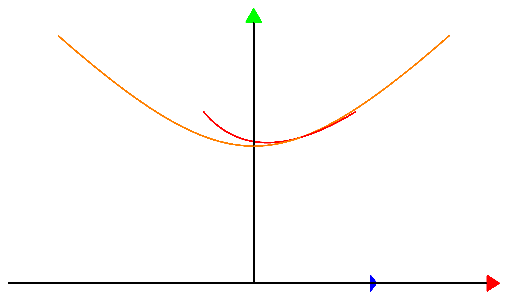}} \qquad
  \subfloat[Hodograph -- oblique view\label{fig:9b}]{\includegraphics[width=4.5cm]{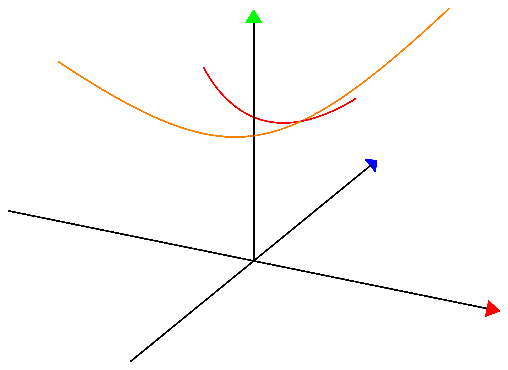}} \\
  \subfloat[Hodograph -- above view\label{fig:9c}]{\includegraphics[width=4.5cm]{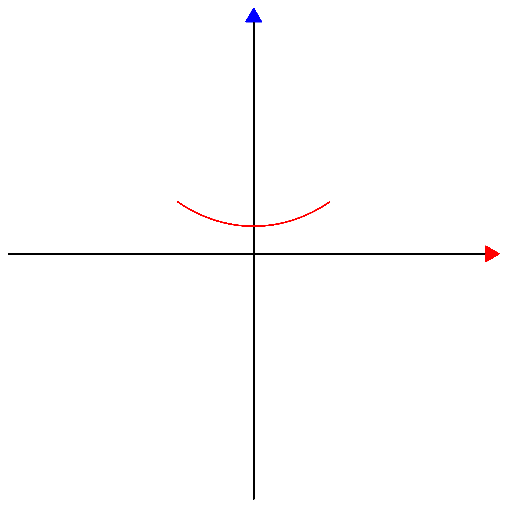}} \qquad
  \subfloat[Spatial orbit \label{fig:9d}]{\includegraphics[width=4.5cm]{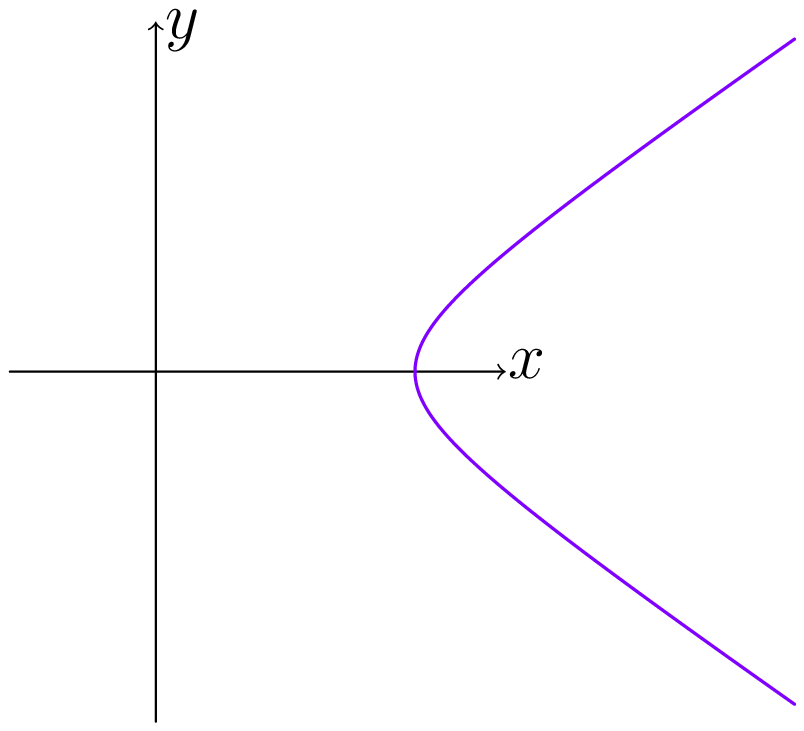}}
  \caption{\label{fig:9}Hodograph and spatial orbit for repulsion for $v_o^\mu$ light-like ($E = 1.25m, \ell = \kappa$).}
\end{figure}

\subsection{Hodographs for $\ell = |\kappa|$} \label{sec: elleq}

Assuming $\kappa = \pm \ell$, the solution of the hodograph equation \eref{eq: uAeq}, compliant with \eref{eq: energyint2} and \eref{eq: urutrel}, is
\begin{equation}\label{eq: ukeql}
\begin{eqalign}
 { \vec u = \pm \frac{E}{m} \left( \theta - \theta_o \right) \hat r \mp \left[ \frac{E}{2m} \left( \theta - \theta_o \right)^2 - \frac{E^2 - m^2}{2mE} \right] \hat \theta \\
 u^0 = \frac{E}{m} \mp u_\theta = \frac{E}{2m} \left(\theta - \theta_o \right)^2 + \frac{E^2 + m^2}{2mE} }
\end{eqalign}
 \end{equation}
with $\theta_o$ an arbitrary shift angle. The condition $u^0 \ge 1$ is clearly satisfied.

\begin{figure}
  \centering
  \subfloat[Hodograph -- side view \label{fig:10a}]{\includegraphics[width=4.5cm]{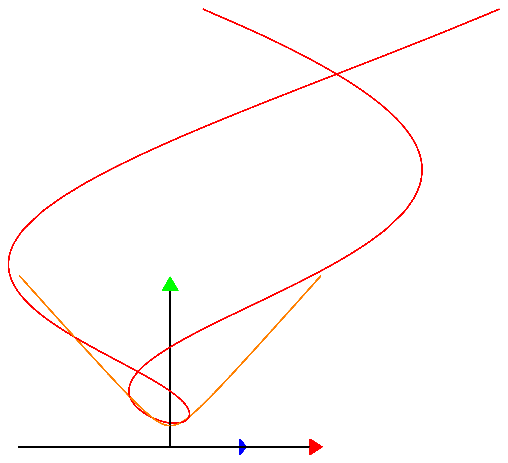}} \qquad
  \subfloat[Hodograph -- oblique view \label{fig:10b}]{\includegraphics[width=4.5cm]{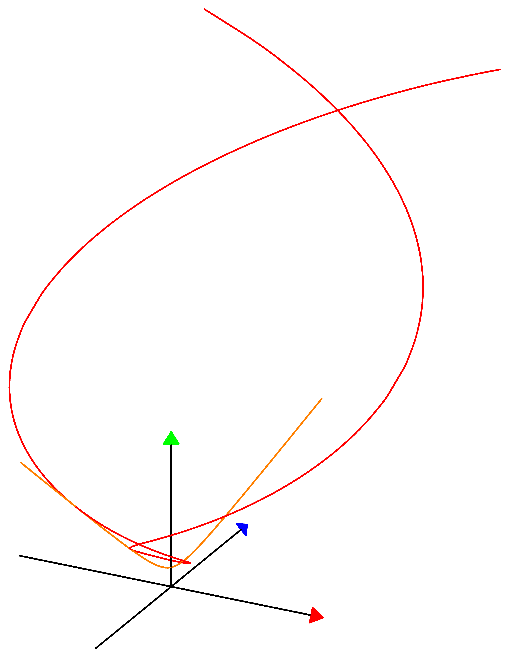}} \\
  \subfloat[Hodograph -- above view \label{fig:10c}]{\includegraphics[width=4.5cm]{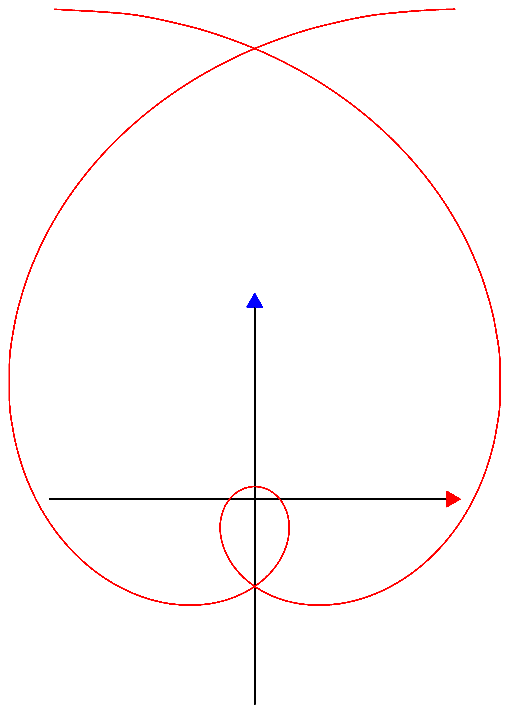}} \qquad
  \subfloat[Spatial orbit \label{fig:10d}]{\includegraphics[width=4.5cm]{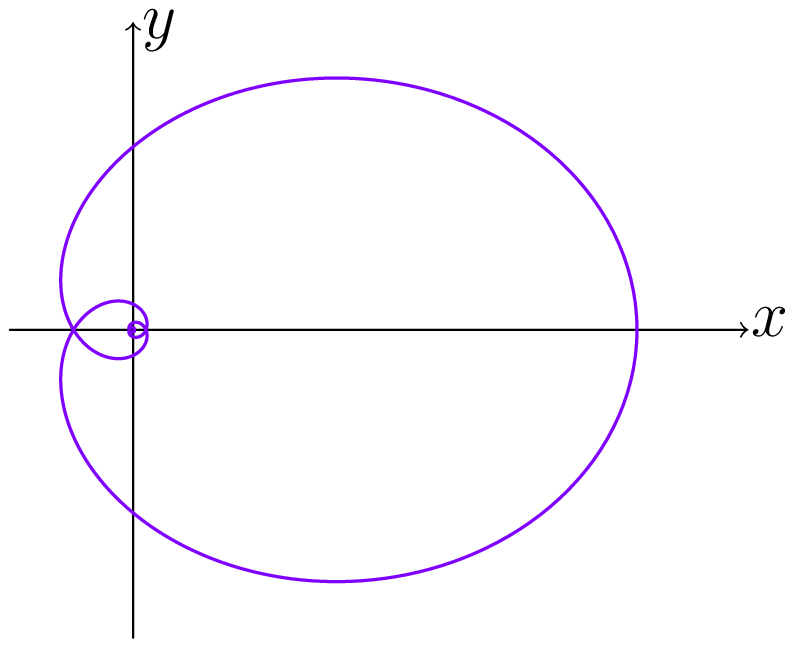}}
  \caption{\label{fig:10} Hodograph and spatial orbit for 1st attraction (unstable bound-like) case for $v_o^\mu$ space-like ($E = 0.6m, \kappa/\ell = -1.05$).}
\end{figure}

\subsection{Structure of the hodographs and trajectories}

The ensuing results are distinguished according to the sign of $\kappa$ :

\begin{figure}
  \centering
  \subfloat[Hodograph -- side view \label{fig:11a}]{\includegraphics[width=4.5cm]{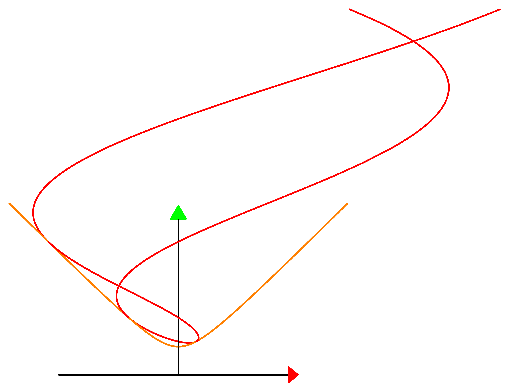}} \qquad
  \subfloat[Hodograph -- oblique view \label{fig:11b}]{\includegraphics[width=4.5cm]{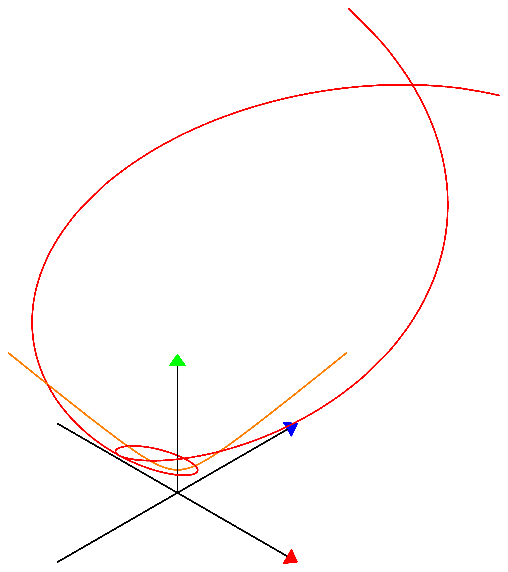}} \\
  \subfloat[Hodograph -- above view \label{fig:11c}]{\includegraphics[width=4.5cm]{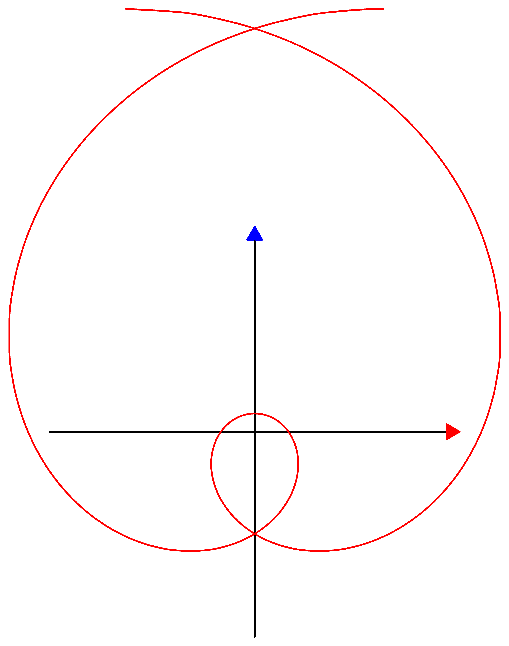}} \qquad
  \subfloat[Spatial orbit \label{fig:11d}]{\includegraphics[width=4.5cm]{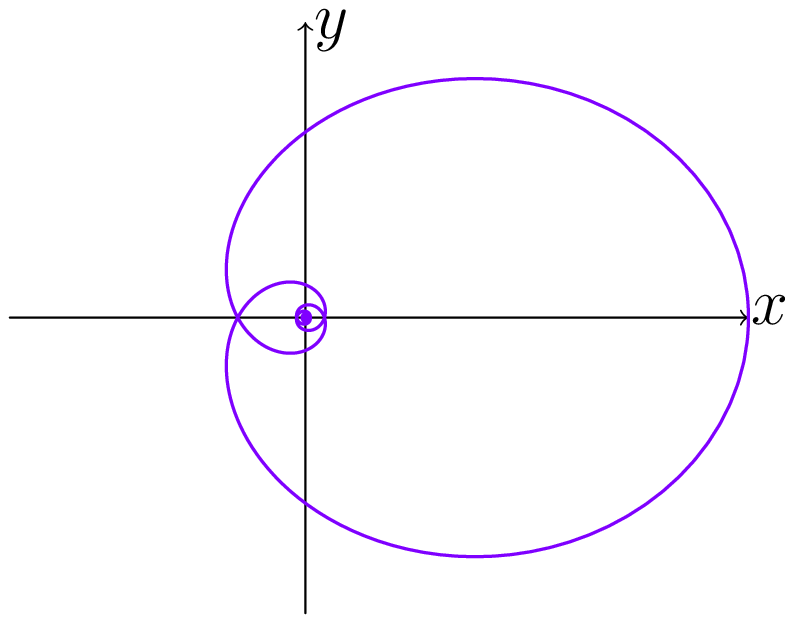}}
  \caption{\label{fig:11} Hodograph and spatial orbit for 1st attraction (unstable bound-like) case for $v_o^\mu$ light-like ($E = 0.6m, \ell = -\kappa$).}
\end{figure}

\subsubsection{Repulsion}

The condition $u_\theta \ge 0$ implies, for $\kappa \ge \ell$, that $\theta$ is bounded in a finite range,
 \begin{equation} \label{eq: utcond-kpos}
 \left| \theta - \theta_o \right| \le \theta_\infty = \cases{
 \frac{1}{\bar \beta} \cosh^{-1} \left( \frac{\kappa E}{m\ell \bar \beta A_o} \right) & $\kappa > \ell$ \\
 \frac{\sqrt {E^2 - m^2}}{E} & $\kappa = \ell$ \\}
 \end{equation}
The condition that $E > m$ for $\kappa \ge \ell$ is verified for both cases in \eref{eq: utcond-kpos}, for the first using the identity
\[
 \left( \frac{\kappa E}{m\ell \bar \beta A_o} \right)^2 - 1 = \frac{1}{{A_o}^2} \left( \frac{E^2}{m^2} - 1 \right) \, .
\]
The particle comes from infinity ($r \rightarrow \infty$, $u_\theta = 0$) to a minimal distance at $\theta = \theta_o$
 \begin{equation} \label{eq: rmin}
 r_{\rm min} = \frac{\ell}{m u_{\theta{\rm max}}} = \cases{
 \frac{\kappa E + m \ell \bar\beta A_o }{E^2 - m^2} & $\kappa > \ell$ \\
 \frac{2\ell E}{E^2 - m^2} & $\kappa = \ell$ \\}
 \end{equation}
and then back again to infinity. Denoting $\theta_{\infty 1} = \theta_o - \theta_\infty$ and $\theta_{\infty 2} = \theta_o + \theta_\infty$, the $\theta$-range is $\theta_{\infty 1} \le \theta \le \theta_{\infty 2}$. As for the Newtonian limit or time-like cases, the infinite trajectory reduces into an hodograph which is a finite segment between $u^\mu_{\infty 1} = \left( E / m, -\left(\sqrt {E^2 - m^2} / m\right) \hat r_{\infty 1} \right)$ at $\theta = \theta_{\infty 1}$ and $u^\mu_{\infty 2} = \left( E / m, \left(\sqrt {E^2 - m^2} / m\right) \hat r_{\infty 2} \right)$ at $\theta = \theta_{\infty 2}$ (see \Fref{fig:8} \& \ref{fig:9}).

\begin{figure}
  \centering
  \subfloat[Hodograph -- side view \label{fig:12a}]{\includegraphics[width=4.5cm]{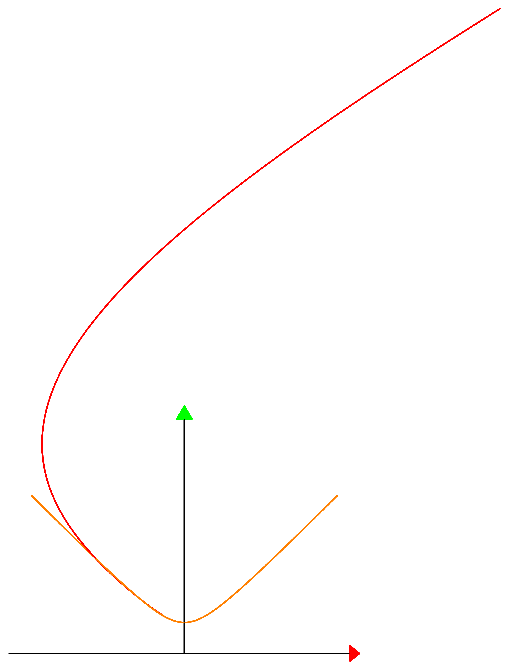}} \qquad
  \subfloat[Hodograph -- oblique view \label{fig:12b}]{\includegraphics[width=4.5cm]{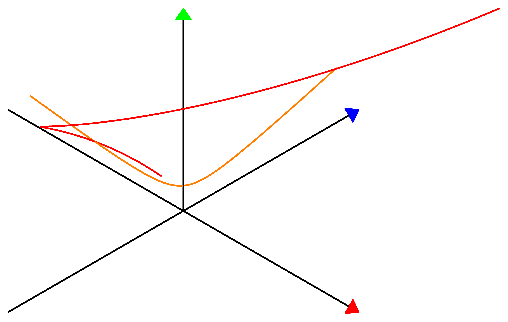}} \\
  \subfloat[Hodograph -- above view \label{fig:12c}]{\includegraphics[width=4.5cm]{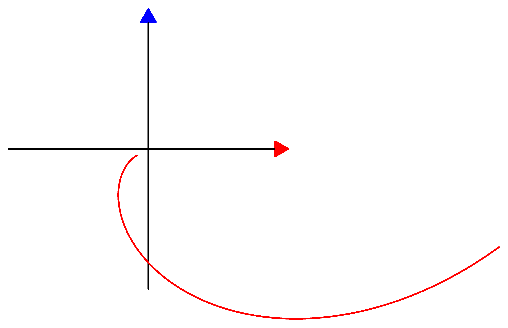}} \qquad
  \subfloat[Spatial orbit \label{fig:12d}]{\includegraphics[width=4.5cm]{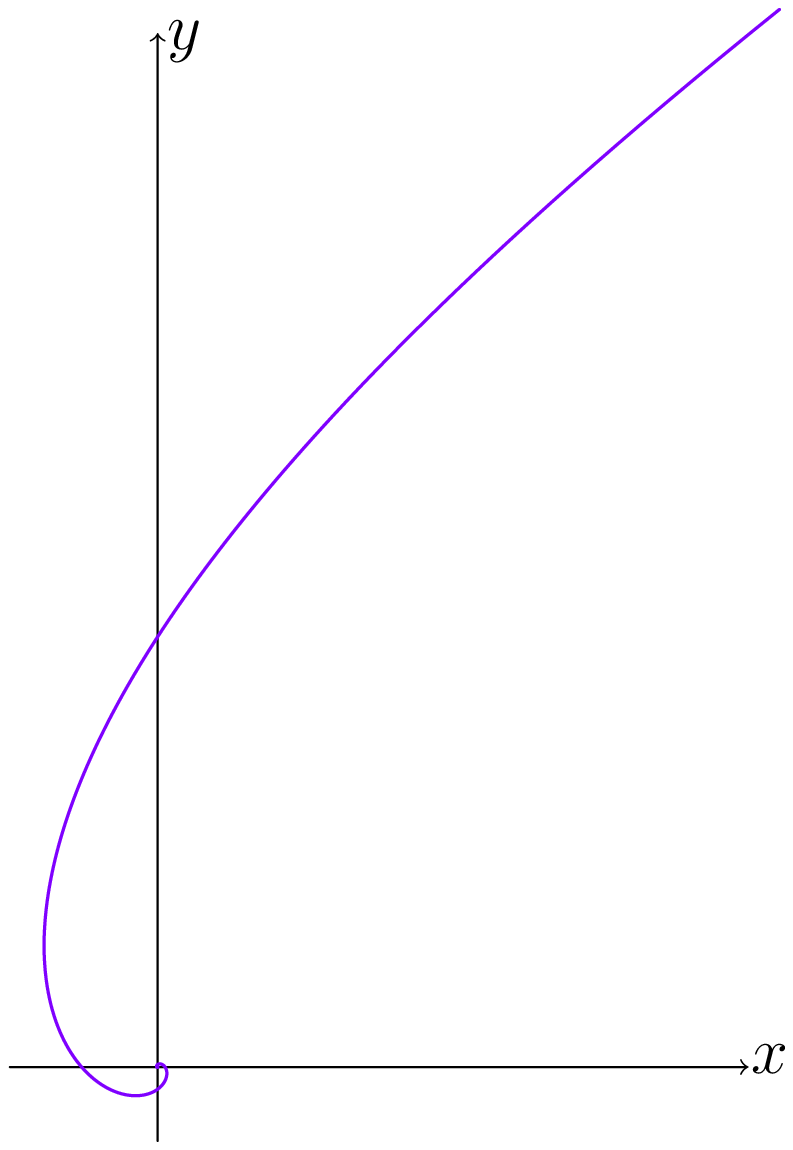}}
  \caption{\label{fig:12} Hodograph and spatial orbit for 2nd attraction (collapse) case for $v_o^\mu$ space-like ($E = 1.25m, \kappa/\ell = -1.2$).}
\end{figure}

\subsubsection{Attraction}

\begin{figure}
  \centering
  \subfloat[Hodograph -- side view \label{fig:13a}]{\includegraphics[width=4.5cm]{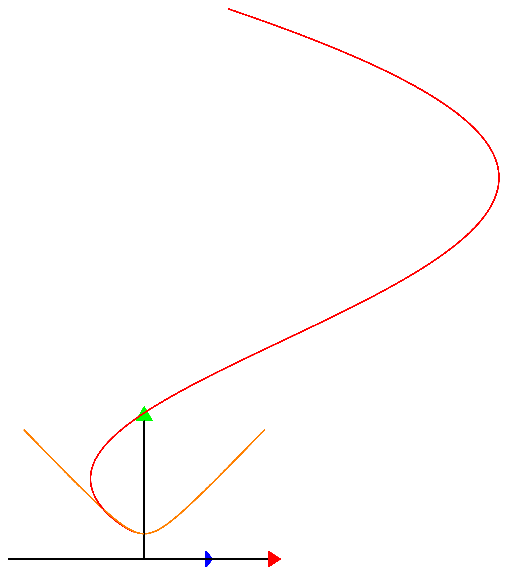}} \qquad
  \subfloat[Hodograph -- oblique view \label{fig:13b}]{\includegraphics[width=4.5cm]{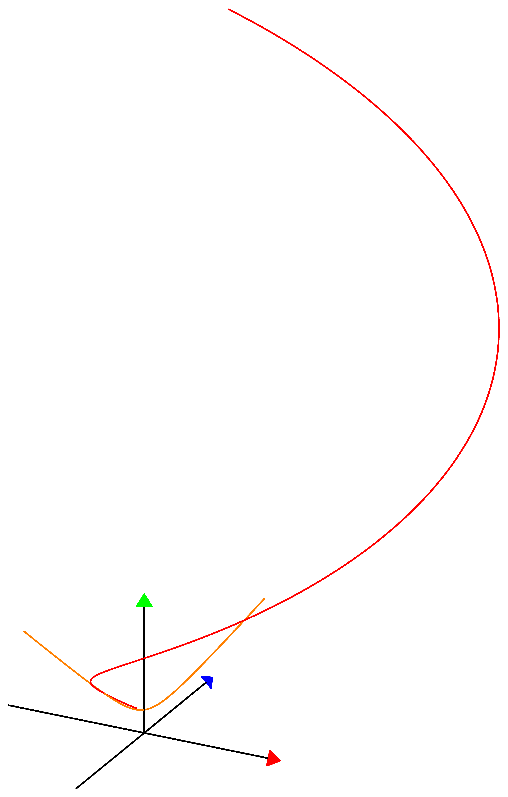}} \\
  \subfloat[Hodograph -- above view \label{fig:13c}]{\includegraphics[width=4.5cm]{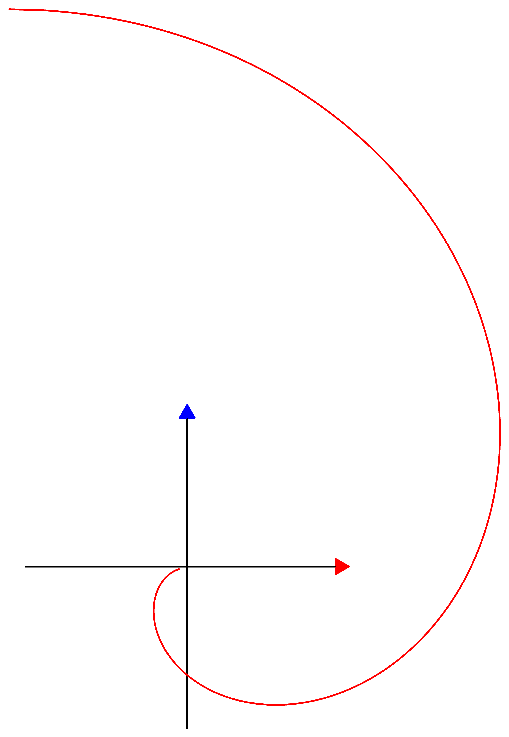}} \qquad
  \subfloat[Spatial orbit \label{fig:13d}]{\includegraphics[width=4.5cm]{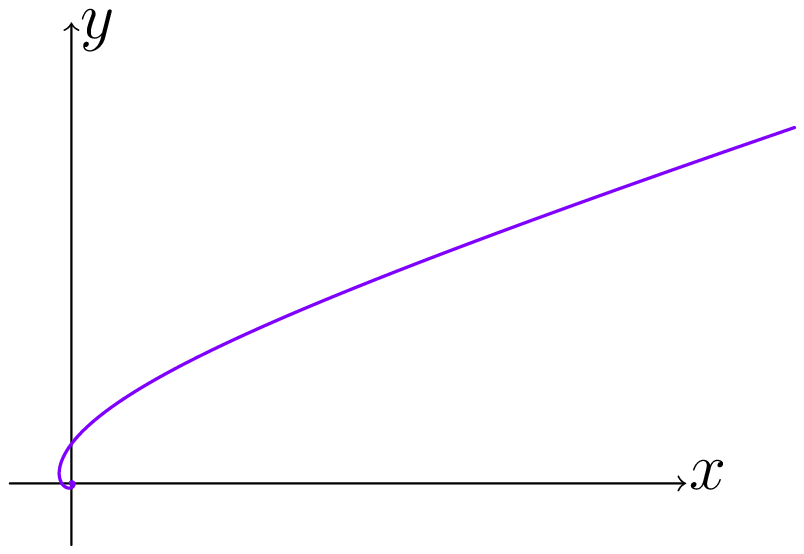}}
  \caption{\label{fig:13} Hodograph and spatial orbit for 2nd attraction (collapse) case for $v_o^\mu$ light-like ($E = 1.05m, \ell = -\kappa$).}
\end{figure}

For $\kappa \le -\ell$ we distinguish between two cases according to $E/m$ :

\begin{enumerate}
 \item
{When $E \ge m$ the condition $u_\theta \ge 0$ implies
 \begin{equation} \label{eq: utcond-kneg}
 \left| \theta - \theta_o \right| \ge \theta_\infty = \cases{
 \frac{1}{\bar\beta} \cosh^{-1} \left( \frac{|\kappa| E}{m\ell \bar\beta A_o} \right) & $\kappa < -\ell$ \\
 \frac{\sqrt {E^2 - m^2}}{E} & $\kappa = -\ell$ \\ }
 \end{equation}
Therefore, a situation similar to gravitational singularity is encountered : Either 1) $\theta_o + \theta_\infty \le \theta < \infty$, then the hodograph ia a semi-infinite segment, the particle arrives from spatial infinity ($u_\theta = 0$) and collapses into the centre of force for $\theta \to \infty$ ($u_\theta \to \infty$, $r \to 0$),

or 2) $ -\infty < \theta \le \theta_o - \theta_\infty$, again a semi-infinite hodograph, and the particle bursts out of the centre of force corresponding to $\theta \to -\infty$ ($u_\theta \to \infty$, $r \to 0$) and escapes to infinity ($u_\theta \to 0$) (although the latter possibility is bizarre and unlikely, it is mentioned because of its theoretical possibility) (see \Fref{fig:10} \& \ref{fig:11}).}
 \item
{When $E < m$, $\theta$ may get any value without bounds, but the particle is spatially bound within the maximum distance from the centre
 \begin{equation} \label{eq: rmax}
 r_{\rm max} = \frac{\ell}{m u_{\theta{\rm max}}} = \cases{
 \frac{|\kappa| E + m \ell \bar\beta A_o }{m^2 - E^2} & $\kappa > \ell$ \\
 \frac{2\ell E}{m^2 - E^2} & $\kappa = \ell$ \\ }
 \end{equation}
The particle may start at some distance from the centre of force, then move in a spiral trajectory towards the centre. At the same time the hodograph starts close to the centre of velocity space and spirals away to infinity. In the limit $\theta \to \infty$ then also $\vec u \to \infty$, and the particle collapses into the centre of force (see \Fref{fig:12} \& \ref{fig:13}).}
\end{enumerate}

\vskip20pt

\section{Concluding remarks} \label{sec: Con}

The purpose of the present work is the application of the generally non-familiar {\it hodograph method} -- studying the dynamics of a system in velocity space -- to a charged relativistic particle in Coulomb field, with the prospects of applying it in the future to general EM 2-body systems.

The hodograph method is applicable, in principle, to all systems with rotational symmetry, translating dynamical systems in ordinary space to geometrical systems in velocity space. It uses the rotational symmetry of the system and the associated angular-momentum conservation to transform spatial dependencies to velocity dependencies and eliminate using the time as evolution parameter (via \eref{eq: uthetar}). The use of rotational symmetry in this way is an essential part of the method. It is important to emphasize that the hodograph method is an analysis confined only to the velocities. The velocity space is the entire arena for the hodograph method, and its geometry plays a major r\^{o}le in the application of the method.

The hodograph method has the merit that when applied to systems with $\kappa/r$ potentials, as the Kepler/Coulomb systems, Newtonian and relativistic alike, the velocity equations are simply linear, providing a very straight-forward and elegant means to analyze the dynamics of the system. These equations and their solutions were discussed and illustrated for the various combinations of the coupling constant $\kappa$ and angular momentum $\ell$. In particular, the deduction of the spatial trajectories from the hodographs was demonstrated.

Perhaps the most significant feature of the relativistic velocity space ${\cal V}_{\rm rel}$ is its being a curved manifold -- hyperboloid $H^3$ -- embedded in the 4-D pseudo-Euclidean space $E^{(1,3)}$. In the Newtonian velocity space, being Euclidean $E^3$, the minimum energy circle  $C_o(\ell)$ retains its shape when translated by the Hamilton vector for higher energy states (\Fref{fig:2} \& \ref{fig:3}). In ${\cal V}_{\rm rel}$, on the other hand, the base circle $C_o(\ell)$, created by the vector $v_o^\mu$ (see \Fref{fig:4}), when shifted necessarily creates a 3-D trajectory. Moreover, it is evident from  \Fref{fig:4} that only in the case $\ell > |\kappa|$ -- for which the spatial velocities on the base circle $C_o(\ell)$ are subluminal -- the vector $v_o^\mu$, being time-like, points towards the interior of $\mathcal{V}_{\rm rel}$. Then, if continued, it punches through the hyperboloid into its interior and draws on it a horizontal circle which is the minimum energy hodograph $u^\mu = \beta^{-1} v_o^\mu(\theta|\ell)$. Only in this case ($\ell > |\kappa|$) a nonrelativistic limit exists. The corresponding spatial trajectories show the familiar features of rotating conic sections \cite{LLF39,Boyer2004}, such as rosettes for bound states or encircling the centre of force in unbound states.

Otherwise, for $\ell \le |\kappa|$, $v_o^\mu$ points outside of the hyperboloid : a light-like $v_o^\mu$ (for $\ell = |\kappa|$), if continued, approaches the hyperboloid asymptotically, while a space-like $v_o^\mu$ (for $\ell < |\kappa|$), if continued, recedes from the hyperboloid. In either case, $v_o^\mu$ cannot draw a circle on the hyperboloid, and no nonrelativistic limit exists. In the absence of a non-relativistic limit the spatial trajectories in these cases feature unfamiliar properties, like spiraling and collapsing into the centre of force \cite{LLF39,Boyer2004}.

The common feature of all Newtonian KC hodographs is them being circles, or circular arcs, whose radius depends very simply on the angular momentum as $R = |\kappa| / \ell$, independently of the energy. The energy dependence enters only through the magnitude of the (constant) Hamilton vector $\vec B_o$, which uniformly displaces the hodograph from the minimum energy state to its actual state; the direction of $\vec B_o$ determines the orientation of the spatial trajectory. The existence of constant or conserved quantity that determines the configuration of a physical system is usually regarded as an indication of a symmetry. It is therefore appropriate to regard these features as constituting a symmetry, which we call {\it Hamilton symmetry} -- a symmetry that acts on the hodographs in velocity space, changing the value $E$ of the energy while keeping the same value $\ell$ for the angular momentum.

Since the Newtonian velocity space is Euclidean, the action of the Newtonian Hamilton symmetry is additive and therefore straight-forward. But on the hyperbolic ${\cal V}_{\rm rel}$ the action of Hamilton symmetry is much more intricate and deserves discussion. The study of Hamilton symmetry for relativistic Coulomb systems is the subject of the companion article \cite{Hamsym}, high-lighting various features of these systems not realized otherwise. Besides the interest in this physical issue \textit{per se}, it is hopes that it may also assist towards a full solution of the relativistic 2-body EM problem.

\vskip20pt

\rule{10cm}{1pt}


 \end{document}